\documentclass[aps,prx,twocolumn,floatfix,nofootinbib,superscriptaddress,longbibliography]{revtex4-2}

\usepackage{graphicx}
\usepackage{multirow}
\usepackage{amsmath,amssymb, amsthm}
\usepackage{mathtools}
\usepackage{soul}
\usepackage{verbatim}

\usepackage{comment}

\usepackage{bm}
\usepackage[colorlinks=true, linkcolor=blue, citecolor=gray]{hyperref}
\usepackage[table]{xcolor}
\usepackage{tikz}
\usepackage{xspace}
\usepackage[capitalize]{cleveref}
\usepackage{import}
\usepackage[caption=false]{subfig}
\usepackage{bbold}
\usepackage[braket, qm]{qcircuit}
\usepackage{chemformula}
\usepackage{adjustbox}
\usepackage{float}
\usepackage{algorithm}
\usepackage{algpseudocode}

\begin{document}

\newcommand{\paragraphsentence}[1]{%
\textbf{\emph{#1}}}
\renewcommand{\paragraphsentence}[2]{#2}

\newcommand{\Google}{\affiliation{%
Google Research, Mountain View, CA 94043, United States}}

\newcommand{\Covestro}{\affiliation{%
Covestro Deutschland AG, 51365 Leverkusen, Germany}}

\newcommand{\Leiden}{\affiliation{%
Instituut-Lorentz, Universiteit Leiden, 2300 RA Leiden, The Netherlands}}

\newcommand{\Quandco}{\affiliation{%
Qu \& Co B.V., 1070 AW Amsterdam, The Netherlands}}

\newcommand{\UMass}{\affiliation{%
Department of Electrical and Computer Engineering, University of Massachusetts Amherst, Amherst MA, USA}}

\newcommand{\UCR}{\affiliation{%
Department of Electrical and Computer Engineering,  University of California, Riverside, CA USA}}

\newcommand{\UCRPA}{\affiliation{%
Department of Physics and Astronomy, University of California, Riverside, CA}}

\newcommand{\UCSB}{\affiliation{%
Department of Physics, University of California, Santa Barbara, CA, USA}}

\newcommand{\Colombia}{\affiliation{%
Department of Chemistry, Columbia University, New York, NY, USA}}

\newcommand{\UConn}{\affiliation{%
Department of Physics, University of Connecticut, Storrs, CT}}

\newcommand{\UAuburn}{\affiliation{%
Department of Electrical and Computer Engineering, Auburn University, Auburn, AL}}

\newcommand{\UTS}{\affiliation{%
QSI, Faculty of Engineering and Information Technology, University of Technology Sydney, NSW, Australia}}

\title{Purification-based quantum error mitigation of pair-correlated electron simulations}

\interfootnotelinepenalty=10000

\author{T.~E.~O'Brien}
\email{teobrien@google.com}
\Google
\Leiden

\author{G.~Anselmetti}
\Covestro

\author{F.~Gkritsis}
\Covestro

\author{V.~E.~Elfving}
\Quandco

\author{S.~Polla}
\Google
\Leiden

\author{W.~J.~Huggins}
\Google

\author{O.~Oumarou}
\Covestro

\author{K.~Kechedzhi}
\Google

\author{D.~Abanin}
\Google

\author{R.~Acharya}
\Google

% ...

\begin{abstract}
An important measure of the development of quantum computing platforms has been the simulation of increasingly complex physical systems.
Prior to fault-tolerant quantum computing, robust error mitigation strategies are necessary to continue this growth. 
Here, we study physical simulation within the seniority-zero electron pairing subspace, which affords both a computational stepping stone to a fully correlated model, and an opportunity to validate recently introduced ``purification-based'' error-mitigation strategies.
We compare the performance of error mitigation based on doubling quantum resources in time (echo verification) or in space (virtual distillation), on up to $20$ qubits of a superconducting qubit quantum processor.
We observe a reduction of error by one to two orders of magnitude below less sophisticated techniques (e.g. post-selection); the gain from error mitigation is seen to increase with the system size.
Employing these error mitigation strategies enables the implementation of the largest variational algorithm for a correlated chemistry system to-date.
Extrapolating performance from these results allows us to estimate minimum requirements for a beyond-classical simulation of electronic structure.
We find that, despite the impressive gains from purification-based error mitigation, significant hardware improvements will be required for classically intractable variational chemistry simulations.
\vspace{0.65cm}
\end{abstract}

\maketitle

The prospect of accurately simulating ground states of quantum systems on quantum hardware has motivated substantial theory and hardware developments over the last decade.
With fault-tolerant quantum computing in its infancy~\cite{Google2022Suppressing} and many years from promised applications~\cite{Reiher2017Elucidating,Burg2021Quantum,Lee2020Even,Goings2020Reliably,Gidney2021How,Campbell2021Early,Berry2022Quantifying} attention has focused on algorithms requiring only short-depth quantum circuits, such as the variational quantum eigensolver (VQE)~\cite{Peruzzo2014Variational,McClean2015Theory,McArdle2020Quantum}.
Theoretical developments in ansatz design~\cite{Wecker2015Progress,Grimsley19Adaptive,Kandala2017Hardware,McClean2015Theory,Elfving2021Simulating,Evangelista2019Exact} and measurement optimization~\cite{Huggins2019Efficient,Cotler2020Quantum,Bonet2020Nearly,Verteletskyi2019Measurement,Crawford2021Efficient,Huang2020Predicting} have enabled small to mid-scale VQE experiments~\cite{OMalley2016Scalable,Kandala2017Hardware,Kandala2019Error,Hempel2018Quantum,Sagastizabal2019Error,Google2020Hartree,Stanisic2021Observing,Kim2021Scalable,Berg2022Probabilistic}.
A key target of variational quantum algorithms has been the electronic structure problem in chemistry~\cite{McClean2015Theory,OMalley2016Scalable,Kandala2017Hardware,Kandala2019Error,Hempel2018Quantum,Sagastizabal2019Error,Google2020Hartree,Motta2022Quantum}.
Such simulations are challenging to implement on quantum hardware due to a long-range two-body fermionic Hamiltonian and stringent accuracy requirements.
This makes it unclear whether a beyond-classical~\cite{Google2019Quantum} simulation of chemistry can be achieved without fault tolerance.
Determining the requirements for such a simulation is a critical open problem.

The electronic structure problem can be expressed in models of varying complexity and realism. Quantum simulations of chemistry within the Hartree-Fock (mean-field) approximation were implemented for system sizes up to $12$ qubits in~\cite{Google2020Hartree}, and this retains the record for the largest VQE calculation of a chemical ground state on quantum hardware.
As a next step, one can consider working in the seniority zero subspace of the entire Hilbert space, which assumes all electrons come in spin-up or spin-down pairs~\cite{surjan2012strongly,kossoski2022hierarchy,gunst2021seniority,Boguslawski2014Efficient,Limacher2013New,Elfving2021Simulating}.
This has the advantage of projecting a local fermionic problem onto a local qubit problem~\cite{Elfving2021Simulating}.
The S$_0$ ground state is not \textit{a priori} classically efficiently simulable~\cite{Elfving2021Simulating} (though good approximate methods are known to exist for many problems~\cite{Dukelsky2003Comment,Dukelsky2012Integrable,Vu2020Size}).
This makes it a good stepping stone beyond Hartree-Fock towards the full electronic structure problem.

Recent quantum experiments have relied on error mitigation techniques~\cite{Cai2022Quantum}, which are not scalable like error correction~\cite{Fowler2012Surface,Google2022Suppressing}, but promise to substantially shrink experimental errors.
Popular methods are based on post-selection~\cite{Mcardle2019Error,Bonet2018Low}, rescaling~\cite{Temme2016Error,Li2017Efficient,Google2020Observation,Montanaro2021Error}, purification~\cite{Huggins2021Virtual,Koczor2021Exponential,Google2020Hartree,Obrien2021Error} and probabilistic cancellation~\cite{Temme2016Error,Endo2017Practical}.
Various schemes and combinations of error mitigation techniques have been implemented in practice~\cite{Sagastizabal2019Error, Kandala2019Error, Google2020Hartree, Google2020Observation, Stanisic2021Observing, Berg2022Probabilistic, Kim2021Scalable, Huo2022Dual}.
However, many of these methods do not promise to remove bias to the level of accuracy needed for useful simulation of chemistry, or remain untested beyond few-qubit experiments.
Shifting from non-interacting fermions to correlated electronic structure, one loses two error mitigation advantages that were crucial to the success of~\cite{Google2020Hartree}: efficient density matrix purification via McWeeny iteration~\cite{McWeeny1960Recent}, and low-cost gradient estimation.

In this work, we mitigate errors accumulated during the preparation of electronic ground states in the seniority-zero space, comparing three different error mitigation techniques --- postselection, echo verification, and virtual distillation --- on up to $20$ qubits of a superconducting quantum processor.
Using either echo verification or a new combination of postselection and virtual distillation, we are able to reproduce the ground state energy and order parameter for an $N=10$-qubit simulation of the Richardson-Gaudin (RG), or pairing model --- the quintessential model of superconductivity --- improving over the unmitigated estimates by $1-2$ orders of magnitude.
This demonstrates an improvement over classical pair-coupled-cluster-doubles, and the non-interacting BCS theory, neither of which are qualitatively correct over the entire range of coupling values considered.
Echo verification was further able to significantly improve over postselected VQE for $6$- and $10$-qubit simulations of the ring-opening of cyclobutene.
While the stringent error requirements ($<0.05$ Hartree) to differentiate between mean-field and the exact solution could only be achieved for the $6$-qubit case, this still represents the largest VQE simulation of electronic structure for chemistry to date.

Finally, we considered the scaling of our simulation of the RG model, using data from simulations at $N=4,6,8,10$.
We observe a clear difference in the asymptotic scaling of the mean absolute error in energy and order parameter when echo verification or virtual distillation are applied.
From this data, we are able to estimate the minimum requirements for a beyond-classical VQE simulation of similar form: a $25\times$ decrease in hardware error rates (from those observed in this work), a limit of $O(N)$-depth for future variational ansatzes, and the need to pre-optimize ansatzes classically without intermediate calls to a device.
Even if this list of requirements is achieved, meeting the high level of accuracy required for the electronic structure problem will pose a serious challenge, as chemical accuracy is around $60\times$ smaller than our mean accuracy for the $10$-qubit cyclobutene problem.

\section{Methods}

\subsection{Simulating the seniority-zero subspace}\label{sec:s0}
The seniority of a Slater determinant is the number of unpaired electrons; thus, the seniority zero (S$_{0}$) sector of Hilbert space for an $N$-electron system in $M$ orbitals is the space of ${M}\choose{N/2}$ determinants leaving no electrons unpaired given a particular pairing of the spin-orbitals.  Seniority is not a global symmetry of the electronic structure Hamiltonian and it is basis dependent; it has been used as a way to classify determinant subspaces to generate better approximations for solving the Schr\"odinger equation~\cite{kossoski2022hierarchy,gunst2021seniority,Boguslawski2014Efficient,Limacher2013New} and as a starting point for modeling strong correlations from electron pair states~\cite{surjan2012strongly}.

Supported by the S$_{0}$ subspace there exists a set of operators satisfying the $su(2)$ algebra constructed from pairs of fermion ladder operators and the spatial orbital number operator~\cite{ring2004nuclear}
\begin{align}
P_{p}^{\dagger} = a_{p\alpha}^{\dagger}a_{p\beta}^{\dagger}  \;\;,\;\; N_{p} = \sum_{p,\sigma}a_{p\sigma}^{\dagger}a_{p\sigma},\nonumber \\
\left[P_{p},P_{q}^{\dagger}\right] = (1 - N_{p})\delta_{p,q} \;\;,\;\;\left[N_{p},P_{q}\right] = -2P_{q}\delta_{p,q},
\end{align}
where $p,q$ and $\alpha,\beta$ are orbital and spin indices respectively.
These operators form a basis for Hamiltonians projected into the S$_{0}$ subspace. The equivalence to an $su(2)$ algebra means seniority zero models resemble Heisenberg spin$-1/2$ models which are easily expressed as Pauli operators.

In this work we focus on two Hamiltonians to validate purification-based error mitigation strategies. The first is the Richardson-Gaudin (RG), or pairing model
\begin{align}\label{eq:main_text_rg}
\hat{H}=& \sum_{p=1}^N \epsilon_{p}N_{p} + g \sum_{p\neq q=1}^N P_{p}^{\dagger}P_{q},
\end{align}
which is a model for a small superconducting grain when $g<0$~\cite{vondelft96parity,braun98superconductivity,dukelsky1999crossover,Scuseria2020Correlating}, but with a $g$-dependent Debye frequency~\cite{braun98superconductivity}.
The second model is the electronic structure Hamiltonian ($H_{\mathrm{elec}}$) projected into the S$_{0}$ subspace
\begin{align}\label{eq:chem_s0}
H_{\mathrm{S}_{0}}&  = P_{\mathrm{S}_{0}}H_{\mathrm{elec}}P_{\mathrm{S}_{0}} 
= \sum_{p}\left(h_{p,p}\right)N_{p} \\
& + \frac{1}{4}\sum_{p\neq q} \left( 2 V_{pqpq}-V_{pqqp}\right)N_{p}N_{q} + \sum_{pq}\left(V_{ppqq}\right)P_{p}^{\dagger}P_{q}. \nonumber
\end{align}
The all-to-all connected Heisenberg spin Hamiltonian is, in general, not known to be classically solvable, but good approximate methods exist.
This is especially true for the RG model, which is often integrable~\cite{Dukelsky2012Integrable}, well-approximated by density-matrix renormalization group~\cite{Dukelsky2003Comment} and pair coupled cluster techniques in the repulsive regime, and solvable by quantum Monte Carlo in the attractive regime (where it has no sign problem).
Pair coupled cluster theory is also known to work well for the electronic structure problem in the S$_0$ subspace~\cite{Boguslawski2014Efficient, Henderson2014Seniority, Stein2014Seniority, Vu2020Size} while full configuration interaction quantum Monte Carlo shows a reduced sign problem~\cite{shepherd2016using}.
As such, although we have strong evidence that the quantum circuits used in this text are not classically simulatable (App.~\ref{app:bqp-completeness}), we do not believe directly scaling  S$_0$ simulations represents the easiest path to a quantum advantage in chemistry; this is instead a stepping stone between a mean field solution and the full electronic structure problem.

\subsection{The unitary pair coupled cluster ansatz and energy estimation}\label{sec:ansatz}

\begin{figure}
    \centering
    \includegraphics[width=\columnwidth]{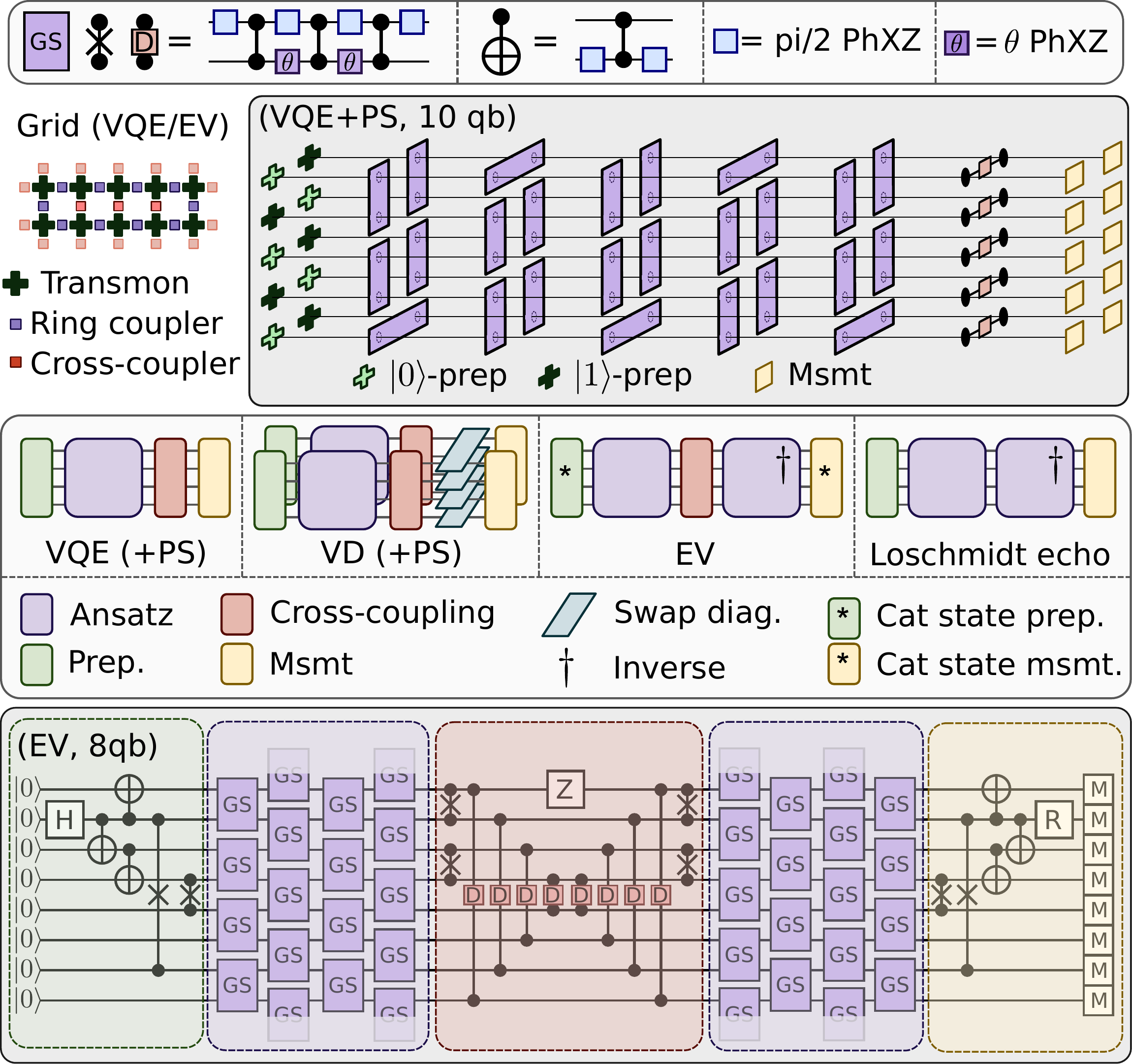}
    \caption{The UpCCD ansatz and its compilation to a 2D superconducting transmon grid. (top) Decomposition of the gates used in this experiment to CZ and single-qubit gates. See supplemental material for details. (second from top, left) $2\times 5$ grid with couplers in a square lattice geometry, showing couplers used during the ansatz (ring coupler, purple), and those used only during measurement (cross-coupler, red). (second from top, right) 2+1D circuit cartoon of a combined ansatz and measurement on a $2\times 5$ transmon qubit array. (third from top) Cartoon of error mitigation techniques used in this experiment. Different circuit pieces are described in the legend. (bottom) an example 8-qubit echo verification circuit to measure the expectation value of $(X_1X_7+Y_1Y_7+Z_1+Z_7)/2$. Shaded gates at the top and bottom of the qubit array wrap around the $2\times 4$ ring.}
    \label{fig:overview}
\end{figure}

In this work we use a Trotterized unitary pair-coupled-cluster doubles (UpCCD) ansatz~\cite{Elfving2021Simulating} compiled into a set of qubits in a ladder geometry with nearest-neighbor coupling.  The ladder ansatz (instead of a generic ring) allows us to efficiently measure terms in the Hamiltonian corresponding qubits that are not physically adjacent after encoding with a minimal number of \textsc{swap} operations. When mapped from fermions to qubits the UpCCD ansatz has the form
\begin{align}
U(\mathbf{\theta}) =& \prod_{\ell}^{N_{\ell}}U_{e}(\theta^{e,\ell})U_{o}(\theta^{o,\ell})\label{eq:UpCCD}\\
U_{o}(\theta^{o,\ell}) =& \prod_{n=0}^{N/2 - 1}GS_{2n+1, (2n+2)\%N}(\theta^{o,l}_{2n+1,(2n+2)\% N}) \\
U_{e}(\theta^{e,\ell}) =& \prod_{n=0}^{N/2 - 1}GS_{2n, 2n+1}(\theta^{e,l}_{2n,2n+1})
\end{align}
where each $GS_{ij}(\theta^{e/o,\ell}_{ij})$ is a Givens-\textsc{swap} gate corresponding to the product of a Givens rotation gate on a pair labeled by qubits $i$ and $j$ followed by a \textsc{\textsc{swap}} operation~\cite{Elfving2021Simulating},
\begin{equation}\label{eq:GSGate}
GS(\phi) = \underbrace{\begin{bmatrix}
1 & 0 & 0 & 0\\
0 & 0 & 1 & 0\\
0 & 1 & 0 & 0\\
0 & 0 & 0 & 1
\end{bmatrix}}_\text{\textsc{swap}}\underbrace{\begin{bmatrix}
1 & 0 & 0 & 0\\
0 & \cos(\phi) & -\sin(\phi) & 0\\
0 &  \sin(\phi) &  \cos(\phi) & 0\\
0 & 0 & 0 & 1
\end{bmatrix}}_\text{Givens}.
\end{equation}
The $GS$ gate corresponds to a coherent partial pair-excitation (by the angle $\phi$), followed by a pair-\textsc{swap}. Given a number of layers $N_{\ell}$ in Eq.~\eqref{eq:UpCCD} and total number of qubits $N$ there are a total of $N_{\ell}N/2$ free parameters in the ansatz. To minimize the amount of time qubits are idle we order the spatial orbitals such that the Fermi-vacuum is $|0101\ldots 01\rangle$--e.g. the restricted Hartree-Fock state--corresponding to an interleaved list of occupied and virtual orbital labels in ascending energy order. The Hamiltonian qubit ordering is then chosen such that when all $\theta=0$, the Hartree-Fock state for each model is returned. The alternating \textsc{swap} gate arrangement allows us to couple each occupied pair with each unoccupied pair once in depth $N/2$ (see App.~\ref{app:scheduling}). Thus, in this work we set $N_{\ell}=N/2$ for all systems. Each $GS(\theta)$ gate is compiled into a product of three controlled-$Z$ ($\textsc{CZ}$) gates interleaved with tunable single-qubit microwave gates (Fig.~\ref{fig:overview} (top), see App.~\ref{app:gates}). 

To perform energy estimation on our two S$_{0}$ models, expectation values with respect to nearest-neighbor and non-nearest-neighbor qubits are required. The expectation value $\langle X_{i}X_{j}+Y_{i}Y_{j}\rangle$ is estimated by performing a number preserving diagonalization~\cite{Google2020Hartree,Bonet2020Nearly} mapping the expectation value to the difference of $\langle Z_{i}\rangle $ and $\langle Z_{j}\rangle$. The ladder geometry allows us to measure all non-nearest-neighbor pairs across the rungs of the ladder in a similar fashion at the additional cost of at most one \textsc{swap} operation. The full measurement protocol is detailed in Appendix~\ref{app:scheduling}. All-to-all coupling is achieved in $N$ circuits bringing the total number of different circuits to measure the Hamiltonian's expectation value to $N+1$.
Strategies with fewer numbers of circuits exist, however they do not allow for post-selection on particle number.

\subsection{Echo verification and virtual distillation}\label{sec:EVVD}

Echo verification (EV), introduced in \cite{Obrien2021Error} is an error mitigation technique that uses two copies of a quantum state $|\psi\rangle$ reflected in time (preparation $\leftrightarrow$ unpreparation) to estimate $\langle\psi|O|\psi\rangle$ for a unitary $O$~\cite{Huo2022Dual,Polla2022Optimizing}.
EV can be implemented without control gates, given a known reference eigenstate $|\phi\rangle$ of $O$ orthogonal to $|\psi\rangle$ (here $|\phi\rangle=|00\ldots\rangle$).
To implement (control-free) EV, we act $O$ on a prepared superposition of $|\psi\rangle$ and $|\phi\rangle$, generated by acting our UpCCD ansatz on the cat state $|00\ldots 0\rangle + |0101\ldots 01\rangle$.
Then, we estimate the expectation value of $|\phi\rangle\langle\psi|$~\footnote{The term $|\phi\rangle\langle\psi|$ is not Hermitian, but may be written as a sum of the Hermitian operators $|\phi\rangle\langle\psi|+|\psi\rangle\langle\phi|$ and $i|\phi\rangle\langle\psi|-i|\psi\rangle\langle\phi|$.} on the resulting state $|\Psi\rangle=O\frac{1}{\sqrt{2}}(|\psi\rangle+|\phi\rangle)$.
The estimation is performed by inverting the preparation unitary.
In the absence of noise, we have
\begin{equation}
    \langle\Psi|\phi\rangle\langle\psi|\Psi\rangle = \tfrac{1}{2}\langle\psi|O|\psi\rangle e^{i\phi},\label{eq:CFEV_expval}
\end{equation}
where $O|\phi\rangle=e^{i\phi}|\phi\rangle$.
The expectation value $\langle\psi|O|\psi\rangle$ can be recovered from Eq.~\eqref{eq:CFEV_expval} as the other terms are known.
The largest effect of noise on the system is to dampen $\langle\Psi|\phi\rangle\langle\psi|\Psi\rangle\rightarrow F \langle\Psi|\phi\rangle\langle\psi|\Psi\rangle$, where $F$ is the circuit fidelity~\cite{Obrien2021Error}.
We can estimate $F$ independently by removing $O$ from the circuit, which yields a Loschmidt echo of the preparation unitary~\cite{Mi2021Information}.
This is achieved in practice by removing a virtual Z rotation (see Fig.~\ref{fig:overview}, bottom), making the estimated Loschmidt fidelity an accurate estimate of $F$.
Further EV implementation details can be found in App.~\ref{app:ev-scheduling}.

Virtual distillation (VD)~\cite{Huggins2021Virtual,Koczor2021Exponential} is an error mitigation technique that uses collective measurements of $k$ copies of a state $\rho$ to estimate expectation values with respect to $\rho^k/\mathrm{Tr}[\rho^k]$.
VD schemes are based on the observation that the cyclic shift operator $S^{(k)}$ is easily diagonalized, and therefore can be measured, which yields e.g. for $k=2$
\begin{equation}
    \mathrm{Tr}[\rho\otimes\rho S^{(2)}]=\mathrm{Tr}[\rho^2],\mathrm{Tr}[\rho\otimes\rho S^{(2)}O_s]=\mathrm{Tr}[\rho^2O],
\end{equation}
with $O_s=\frac{1}{2}(I\otimes O + O\otimes I)$.
$S^{(2)}$ can be simultaneously diagonalized with $O_s$ when $O=Z_i$ by a $GS(\pi/4)$ rotation between pairs of identified qubits on the two registers.
For two $N/2\times 2$ ladders on a square lattice geometry, this requires one round of \textsc{swap} gates to shift identified qubits next to each other.
Operators $O\neq Z_i$ are measured by rotating to $Z_i$ (see Sec.~\ref{sec:ansatz}) and following the above procedure.
The virtual distillation circuit is only $6$ two-qubit gates deeper than post-selected VQE.

As the $GS(\pi/4)$ gate is number-conserving, VD can be combined with postselection: the global excitation number $\sum_j(Z_j\otimes I+I\otimes Z_j)$ is a good symmetry.
This requires that the state prior to measurement also conserve number.
This is true when estimating $\langle X_iX_j+Y_iY_j\rangle$, but not when estimating $\langle Z_iZ_j\rangle$: when mapping $Z_iZ_j\rightarrow Z_i$ one can only preserve the parity of the total number of excitations.
In the main text of this work, we will present results showing VD with postselection only (PS-VD).
We compare VD with and without postselection in App.~\ref{app:VD-PS-comparison}.

\section{The Richardson-Gaudin model}

\begin{figure}
    \centering
    \includegraphics[width=1\columnwidth]{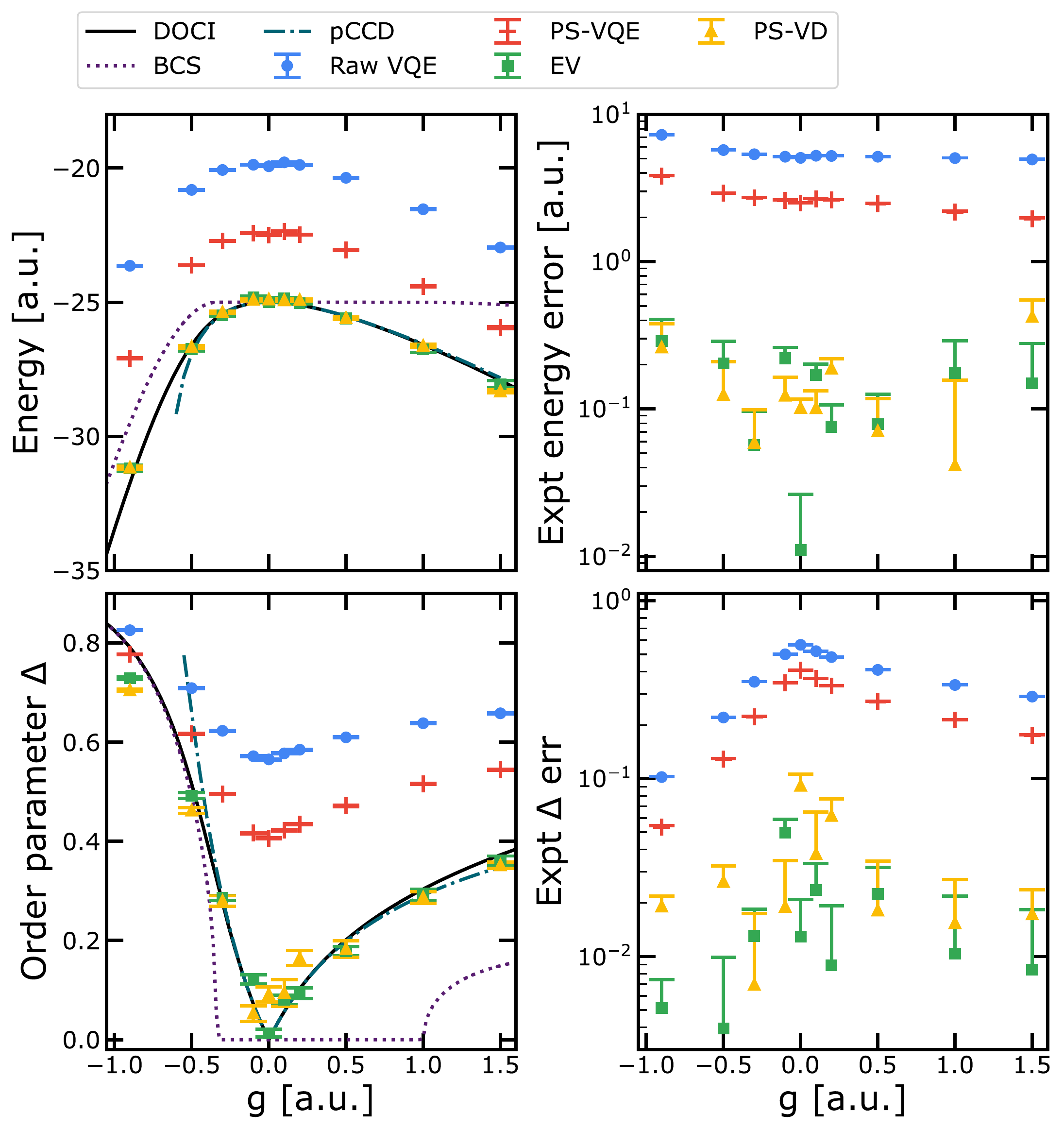}
\caption{Digital quantum simulation of ground states of the RG Hamiltonian for $10$ spatial orbitals (Eq.~\eqref{eq:main_text_rg} and Eq.~\eqref{eq:rg_chem_pot}) on a superconducting quantum device. (top-left) Energy as a function of the coupling parameter $g$, for an unmitigated state preparation [blue circles], and state preparation mitigated by postselection [red crosses], echo verification [yellow triangles], and postselected virtual distillation [green squares]. This is compared to the exact DOCI result [black solid line], and BCS [purple dashed line] and pCCD [teal dashed-dotted line] classical approximations. The pCCD results do not converge below a critical value, resulting in their cut-off. (top-right) Log plot of experimental energy error (ignoring the model error from the UpCCD approximation). (bottom-left) Many-body order parameter for the RG Hamiltonian (see text), again compared to classical models. (bottom-right) Experimental error in estimating the superconducting order parameter vs the target state within the UpCCD approximation (again ignoring model error). Standard deviation error bars estimated by propagating variance (Raw VQE, PS-VQE) or bootstrapping (EV, PS-VD), see App.~\ref{app:errorbars} for details.}
    \label{fig:RG_figure}
\end{figure}

We use our UpCCD ansatz to prepare approximate ground states of the RG model on 10 sites at half-filling across a range of coupling strengths using parameters optimized in noiseless simulations. We achieve half-filling by adding a chemical potential to the single particle energies in Eq.~\eqref{eq:main_text_rg};
\begin{equation}\label{eq:rg_chem_pot}
    \epsilon_p = p - \mu, \hspace{0.5cm} \mu=\frac{1}{2}(N+1)
\end{equation}
In Fig.~\ref{fig:RG_figure} (top left), we estimate the prepared states' energy with and without error mitigation techniques (see caption), and compare it to exact diagonalization in the S$_{0}$ subspace, also known as double occupied configuration interaction (DOCI), and classical pair-coupled-cluster doubles (pCCD), and BCS solutions.
We see that using EV or PS-VD we are able to reproduce the entire energy curve to high accuracy, which neither pCCD nor the non-interacting BCS theory can achieve.
The experimental error in the result is the sum of the UpCCD model error and the experimental error.
To disambiguate the effects of UpCCD model error, in Fig.~\ref{fig:RG_figure} (top right) we plot the error between our experimental data and the UpCCD ground state energy.
Postselection consistently mitigated around half the error present in the raw ansatz.
By contrast, EV demonstrates an average $85$-fold and maximum $460$-fold error reduction.
PS-VD achieves similar performance, with an average $60$-fold and maximum $140$-fold improvement.
The residual error following EV or PS-VD drifts notably with fluctuations between points larger than error bars.
We attribute this observation to device drift.

The RG Hamiltonian has a well-known phase transition in the attractive regime ($g\leq 0$) in the thermodynamic limit, which appears in the BCS state at finite $N$, but is not present in the true ground state due to finite size effects~\cite{vondelft96parity,braun98superconductivity,dukelsky1999crossover}.
This presents an opportunity for a variational quantum simulation to determine qualitative features of a quantum Hamiltonian beyond non-interacting physics.
The traditional order parameter for the BCS state, $\Delta_{BCS}=\frac{1}{N}\sum_j\langle a_{j\uparrow}a_{j\downarrow}\rangle$, is zero on the RG Hamiltonian ground state due to number conservation.
However, one can confirm that $\Delta = \frac{1}{N}\sum_{j,\sigma}\sqrt{\langle n_{j\sigma}^{2}\rangle-\langle n_{j\sigma}\rangle^2}$ satisfies $\Delta=\Delta_{BCS}$ for the BCS ground state of the Hamiltonian, giving a many-body order parameter~\cite{braun98superconductivity}.
In Fig.~\ref{fig:RG_figure} (bottom left), we plot experimental estimates of $\Delta$ across the range of $g$ values considered.
In the absence of error mitigation, though the order parameter dips around $g=0$ the true cusp is not reproduced.
Both EV and PS-VD clearly improve over the BCS approximation for $g>0.5$, with EV particularly able to reproduce the cusp at $g=0$.
The performance of error mitigation is demonstrated by plotting the error in $\Delta$ against the noise free UpCCD energy in Fig.~\ref{fig:RG_figure} (bottom right).

We see all experimental estimates have a slight peak in error at $g=0$.
This can be attributed to $\Delta$ being highly sensitive to error at this point ($\frac{\partial\Delta}{\partial\langle n_{j\sigma}\rangle}\rightarrow\infty$).
Furthermore, the maximally-mixed state has $\Delta=1$, so decoherence has a larger effect when targeting $\Delta<<1$.
This contrasts with the error in the energy (Fig.~\ref{fig:RG_figure} (top right)), which has a slight dip near $g=0$.
We attributed this to the increased contribution from $\langle X_iX_j+Y_iY_j\rangle$ to the energy when $g$ is far from $0$, as $\Delta$ is independent of these expectation values.
The improvement from EV and PS-VD in estimating the order parameter is slightly less than that in estimating the energy, with a mean (max) $32$-fold ($56$-fold) improvement from EV, and $18$-fold ($51$-fold) improvement from PS-VD.
We attribute this to the increased sensitivity of $\Delta$ to noise at $g=0$, and the high performance of the raw results at $g<<0$ (where depolarizing noise has little effect as $\Delta\sim 1$).

\begin{figure}[ht!]
    \centering
    \includegraphics[width=1\columnwidth]{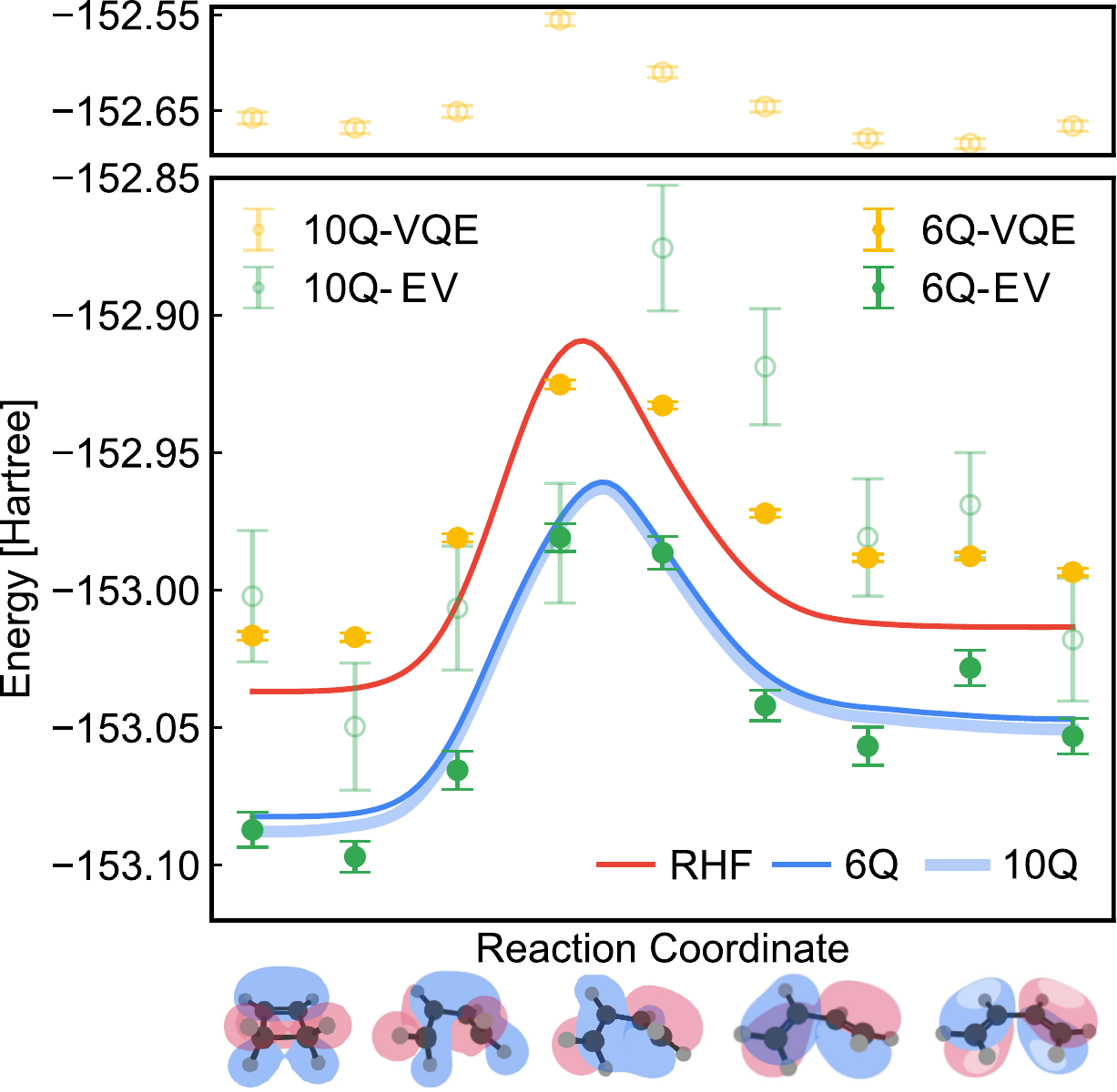}
    \caption{The conrotatory Cyclobutene ring opening pathway simulated in the seniority zero subspace comparing post-selected VQE and echo verification (EV) on an optimized unitary pair-coupled-cluster ansatz. From left to right the reaction path corresponds to the ring opening reaction. For the ten orbital case the unitary pair-coupled-cluster ansatz (evaluated in simulation) has less than 1.8$\times 10^{-4}$ energy difference from exact diagonalization in the seniority zero space. The blue curves correspond to the exact diagonalization of the seniority zero active space Hamiltonian spanning 10 orbitals (lighter-broad blue line) and 6 orbitals (darker-narrow blue line). The red curve is the restricted Hartree Fock (RHF) mean-field energy. Green points (darker green for 6 qubits and lighter green for 10 qubits) are the echo verified experimental data while yellow points (darker yellow for 6 qubits and lighter yellow for 10 qubits) are the post-selected VQE energies.  The 10 qubit VQE data is plotted on a discontinuous and different scale to preserve the visual scale of the reaction energy along the reaction coordinate.} 
    \label{fig:cyclobutene_fig}
\end{figure}

\section{Cyclobutene ring opening}

We further validated scalable error mitigation protocols by simulating the conrotatory ring opening pathway for cyclobutene in an active space of six orbital and six electrons and ten orbitals and ten electrons corresponding to a six and ten qubit simulation of the Hamiltonian in Eq.~\eqref{eq:chem_s0}.  The mechanism of this ring opening is described by the Woodward-Hoffmann rules for pericyclic ring openings corresponding to the in-phase combination of the two carbon $2p$ orbitals when brought together to form the four-member carbon ring.

The geometries along the reaction path are determined from a nudged elastic band calculation using density functional theory (B3LYP) to evaluate forces. The final structures use a minimal basis set (STO-3G) to generate the active space Hamiltonians to project into the seniority zero sector. The Woodward-Hoffmann rules are a type of molecular orbital theory and thus we expect this reaction to be qualitatively described within mean-field theory. This is verified numerically for our seniority zero model where the largest CI coefficient has an average value of 0.974(9), for six-orbitals, and 0.973(9), for 10-orbitals, indicating a single-reference system.  As such, our unitary pair-coupled-cluster doubles ansatz targets the dynamic correlation corrections to the mean-field. 

The average post-selected-VQE absolute error is $0.058 \pm 0.006$ and $0.395 \pm 0.023$ Hartree for the six orbital and ten orbital systems, respectively. The average echo-verified absolute error is $0.011 \pm 0.005$ and $0.064 \pm 0.035$ Hartree for the six orbital and ten orbital system, respectively, showing a 5.51-fold and 6.12-fold improvement over post-selected-VQE average error. Comparing to the raw VQE data, we find a 55.1-fold and 38.4-fold mean error reduction for the six orbital system and 10-qubit system respectively. While there is notable improvement in energy across the reaction pathway for the 10 orbital system the magnitude of the errors is larger than the 0.037 Hartree energy difference between cyclobutene and 1,3-Butadiene. Furthermore, a visual inspection of Figure~\ref{fig:cyclobutene_fig} indicates high parallelity errors in the 10 orbital system. Given the error bars on echo verification are smaller than the parallelity error (point scatter) we attribute the main source of error to device drift.

\section{Outlook}\label{sec:scaling}

\begin{figure}[t]
    \centering
    \includegraphics[width=\columnwidth]{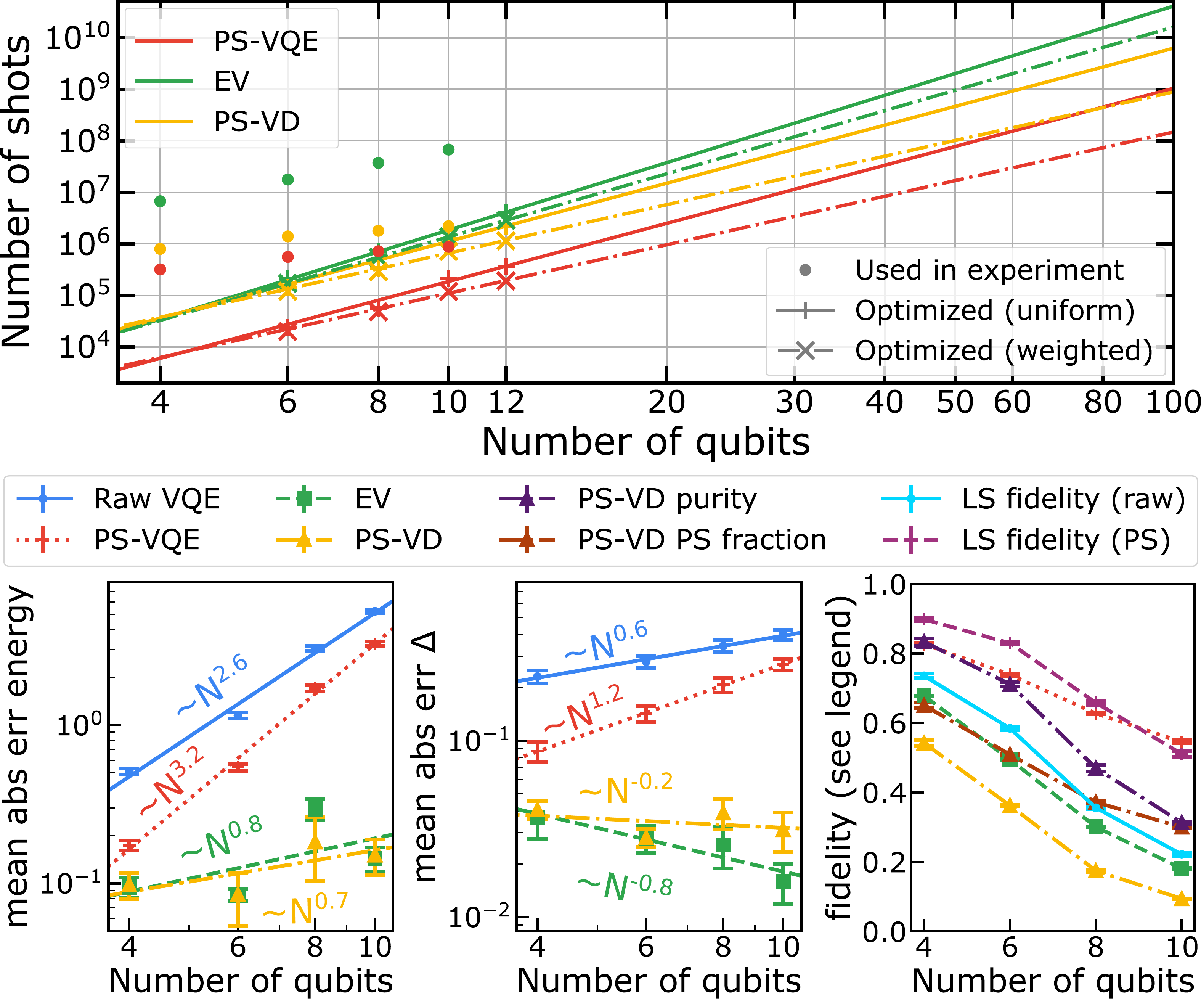}
    \caption{Scaling the simulation of the RG model to larger qubit counts. (top) Number of shots required for convergence at $g=-0.9$. Dots give numbers chosen for the experiment, crosses and pluses give simulated estimations using two types of term grouping (see App.~\ref{app:shot_distribution_expt}) using observed fidelities of a $10$ qubit experiment. (bottom-left) Experimental energy error (vs the UpCCD ground state), averaged over all points studied of the RG model. Error bars show sample standard deviation, and lines a power-law fit (exponent shown) as a guide to the eye. (bottom-middle) Experimental error in order parameter (vs the UpCCD ground state), averaged over all points studied of the RG model. Error bars and lines same as bottom-left. (bottom-right) Different fidelity metrics for post-selected VQE, EV, VD, and Loschmidt echo (see legend), averaged over all points studied of the RG model.}
    \label{fig:scaling_figure}
\end{figure}

We have observed the echo verification and virtual distillation error mitigation protocols suppressing errors by 1-2 orders of magnitude on a range of quantum simulation problems using up to 20 superconducting qubits. We now consider the requirements for scaling these experiments to the classical intractable regime.

In Fig.~\ref{fig:scaling_figure} (top) we plot the number of experiments (shots) used in this work to simulate the RG Hamiltonian at $g=-0.9$ (where pCCD does not describe the system well), and compare this to theoretical estimates targeting the same model to within a sampling noise of $0.1~\mathrm{a.u.}$ using the experimental fidelities observed for $10$ qubits
(fidelities taken from Fig.~\ref{fig:scaling_figure} (bottom right)).
The $50\times$ gap between theory and experiment for $10$-qubit EV can be attributed mostly to extra circuits used to cancel out a background magnetic field (see App.~\ref{app:ev-scheduling}).
The gap for our VD experiment is roughly $3\times$ by comparison.
Assuming the ability to freely weight our shot distribution, we estimate that for a 50-qubit experiment (as a proxy lower bound for a beyond-classical quantum computation) using VD or EV, $10^8$ or $10^{9}$ shots would be required respectively.
This is executable on current hardware in a wall-clock time (see App.~\ref{app:wall-clock_time_model}) of $>1$ hour or $>10$ hours respectively.
Including the difference between experiment and theory at $10$ qubits raises the cost of EV to $5\times 10^{10}$ shots, which would require multiple days to achieve.
These numbers do not include the multiplicative cost of variational optimization (see App.~\ref{app:optimization}).
Furthermore, the requirements for accurate electronic structure simulations may be lower than the $0.1~\mathrm{a.u.}$ requirement considered here.
Methods to pre-optimize variational ansatzes classically, and applications of VQE to problems simpler than electronic structure, may thus be necessary for beyond-classical VQE experiments.

Device coherence presents an additional scaling challenge.
To maintain circuit fidelity $F$ over an $O(N)$-depth, fully parallel circuit as $N$ scales from $5$ to $50$ requires all error rates to drop by and coherence times to increase by roughly a factor $25$ (proportional to $O(N^2)$).
As any reduction in $F$ incurs an $O(\mathrm{poly}(F^{-1}))$ sampling cost~\cite{Obrien2021Error,Huggins2021Virtual}, and as $F$ scales exponentially in the error rate, and as $F\sim 10\%$ for PS-VD (Fig.~\ref{fig:scaling_figure} (bottom-right)), we see little room for negotiation on this $25\times$ lower bound.
To achieve a $25\times$ decrease in error rate would require a $50$-qubit device with XEB fidelities on all two-qubit gates $\leq 3\times 10^{-4}$.
However this analysis precludes ansatzes with depth $O(N^2)$ or higher or significantly larger constant factors (in our case, the circuit depth of the bare VQE is $3N/2$).
For instance, successfully implementing a $50$-qubit VQE with ansatz depth $3N^2/2$ with EV or VD would require error rates to drop $\sim 1000\times$.

On a more positive note, in Fig.~\ref{fig:scaling_figure} (bottom left), we plot the absolute error in the energy estimates, averaged across all points in our RG model experiment.
The energy scales sublinearly after applying EV or VD (a clear asymptotic difference to raw or postselected VQE), which suggests that a $25\times$ decrease in error rate required to keep sampling costs constant may yield significantly higher precision results.
A similar gap between EV/VD and VQE/PS-VQE for estimating the order parameter can be observed in Fig.~\ref{fig:scaling_figure} (bottom middle); the discrepancy in absolute scaling can be attributed mostly to the energy scaling as $O(N^2)$, while $\Delta$ does not scale with $N$.
This observation runs contrary to the observations in Fig.~\ref{fig:cyclobutene_fig}, where shifting from $6$ to $10$ qubits increased the mean error by a factor $10$.
Investigating the mean error in estimating Pauli operators (App.~\ref{app:pauli_error_histogram}) suggests that the true scaling lies somewhere in between these values.
If the energy error scales linearly or better with the error rate per qubit (which is expected from simulations in Ref.~\cite{Obrien2021Error}), and scales less than quadratically in $N$, our requirement to scale error rates as $O(N^{-2})$ to preserve the circuit fidelity $F$ will yield a drop in absolute energy error as a function of $N$.
Thus, pinning down this scaling of experimental error with system size and error rates is a key area for future work.

\subsection*{Author contributions}
T.E.O. calibrated the device and ran the experiments. V.E.E., G.A., and C.G. designed the UpCCD ansatz. V.E.E., G.A., and T.E.O. designed the scheduling of the ansatz onto a $2\times N$ grid. F.G. and C.G. designed the conjugate model gradient descent algorithm and pre-optimized ansatz parameters for the ansatz. N.C.R. wrote the pair coupled cluster code, pre-optimized ansatz parameters, and performed the classical chemistry calculations for the cyclobutene model. T.E.O., W.J.H., S.P., K.K. and R.B. designed and optimized the error mitigation and EV measurement strategies. O.O. and C.G. developed the BQP completeness proof for the UpCCD ansatz. T.E.O., N.C.R., C.G., F.G., V.E.E. and R.B. wrote the paper. T.E.O., C.G., R.B. and N.C. led and co-ordinated the project. All authors contributed to revising the manuscript and writing the Supplementary Information. All authors contributed to the experimental and theoretical infrastructure to enable the experiment.

\subsection*{Acknowledgments}
Some discussion and collaboration on this project occurred while using facilities at the Kavli Institute for Theoretical Physics, supported in part by the National Science Foundation under Grant No. NSF PHY-1748958.

\subsection*{Data availability}
Raw and processed experimental data can be found at https://doi.org/10.5281/zenodo.7225821 

\bibliography{vpe.bib}

\appendix
\section{Calibration of the processor}\label{sec:insitu}
All experiments were implemented on a subgrid of a 25-qubit superconducting processor with the Sycamore architecture.
For all methods other than virtual distillation, a $2\times N/2$ qubit grid was calibrated to within $0.008$ XEB fidelity~\cite{Google2019Quantum} and $0.008$ speckle purity~\cite{Google2019Quantum}.
For virtual distillation, a $4\times N/2$ qubit grid was calibrated to within $0.01$ XEB fidelity and $0.01$ speckle purity.

We were further required to calibrate the single-qubit Z-phases accumulated during a CZ gate.
This is a well-documented issue~\cite{Neill2021Accurately, Google2020Observation}, but is complicated in our case by the addition of microwave gates.
These are observed to bleed into the CZ gate, which made standard Floquet calibration techniques inaccurate.
To solve this issue, we calibrate CZ gates \emph{in-situ}.
The Givens-\textsc{swap} gate was altered by, after each CZ between qubits $i$ and $j$, inserting virtual rotations $\exp(iZ_i\beta_i^{(j)}),\exp(iZ_j\beta_{j}^{(i)})$ on qubit $i$ and $j$ respectively.
The phases $\beta_i^{(j)}$ were calibrated by running two experiments in series.
Firstly, a single $GS(0)=\textsc{swap}$ gate was implemented between qubits $i$ and $j$ (with virtual gates inserted); the qubits were prepared in the state $|0+\rangle$ measured in the $ZX$ or $ZY$ basis, or prepared in the state $|+0\rangle$ and measured in the $XZ$ or $YZ$ basis.
Sweeping $\beta_i^{(j)}$ and $\beta_j^{(i)}$ gave four datasets that could be fitted to extract optimal phase offsets.
The resulting gate was then benchmarked by estimating $\langle XI\rangle$ and $\langle YI\rangle$ on the state $[GS(0)]^{2k}|0+\rangle$ and $\langle IX\rangle$ and $\langle IY\rangle$ on $[GS(0)]^{2k}|+0\rangle$, and fitting this to an oscillatory decay curve.
Under this benchmark, the initial calibration typically reduced the accumulated phase per CZ to less than $30$ milliradians.
This benchmark was further used to calibrate, by sweeping $\beta_i+\beta_j$ on pairs $i,j$ that are being acted on by the same GS gate to remove the remaining oscillations.
We find in practice that a cubic fit to $11$ datasets is a robust way to perform a final estimate of $\beta_i+\beta_j$, with the residual phase less than $5$ milliradians when calibration was successful.
If the estimated fidelity of the resulting GS gate underperformed ($>1.5\%$ error per CZ gate), qubit or coupler frequencies were reoptimized before recalibrating.
Calibration was performed in parallel on sets of CZ gates that were run in parallel during an experiment, to mimic the local environment and compensate for 2-qubit gate crosstalk.

\section{Further details of the UpCCD ansatz}

\subsection{BQP-completeness of nearest neighbor Givens-\textsc{swap} circuits}
\label{app:bqp-completeness}
Here we substantiate the claim that the UpCCD circuits realized on hardware in this work are in general not efficiently classically simulable.
We do so by constructing a universal quantum gate set on a reduced Hilbert space (dual-rail encoding) with an $O(1)$ depth overhead.
This construction shows that any nearest neighbor depth-$O(N)$ circuit on a line of qubits can be mapped to a depth-$O(N)$ UpCCD ansatz (and circuits with arbitrary connectivity to a depth-$O(N^2)$ UpCCD ansatz), when allowing for the omission of gates (as the identity is not a GS gate).
For this to hold it is pivotal that the $GS$ gate family includes the \textsc{\textsc{swap}} gate and is thus not a matchgate.

To demonstrate a universal gate set we use a dual-rail encoding of one logical qubit into two physical qubits (onto which the $GS$ gates will act).
We use tilde to denote logical states and operations and set 
\begin{align}
    | \Tilde{0} \rangle &\coloneqq | 01 \rangle \\
    | \Tilde{1} \rangle &\coloneqq | 10 \rangle .
\end{align}
It is then straightforward to verify by direct computation that a $GS$ gate acting on two physical qubits belonging to the same logical qubit can be used to realize the following logical Hadamard, Pauli, and Pauli rotation gates:
\begin{align}
    GS(\frac{\pi}{4}) &= \widetilde{H} \\
    GS(0) &= \widetilde{X} \\
    GS(\frac{\pi}{2}) &= \widetilde{Z} \\
    GS(0) \cdot GS(\theta) &= \widetilde{RY}(\theta) 
\end{align}
Logical two qubit entangling gates can be realized by acting with $GS$ gates on qubits belonging to two different logical qubits.
The $\widetilde{\textsc{cnot}}$ gate can for instance be made by means of
\begin{equation}
\begin{array}{cc}
\Qcircuit @R=1em @C=0.75em {
 & \multigate{1}{\widetilde{X}} & \qw & \qw & \qw & \multigate{1}{\widetilde{X}} & \qw & \multigate{1}{\widetilde{Z}} & \qw \\
 & \ghost{\widetilde{X}} & \qw &\multigate{1}{\widetilde{Z}}& \qw & \ghost{\widetilde{X}} & \qw  & \ghost{\widetilde{Z}} & \qw \\ 
 & \multigate{1}{\widetilde{H}} & \qw & \ghost{\widetilde{Z}}& \qw & \multigate{1}{\widetilde{H}} & \qw & \qw & \qw \\
 & \ghost{\widetilde{H}} & \qw & \qw & \qw & \ghost{\widetilde{H}} & \qw & \qw & \qw \\
}
\end{array}
=
\begin{array}{c}
\Qcircuit @R=1em @C=0.75em {
 & \multigate{3}{\widetilde{\textsc{cnot}}} & \qw \\
 & \ghost{\widetilde{\textsc{cnot}}} & \qw \\
 & \ghost{\widetilde{\textsc{cnot}}} & \qw \\
 & \ghost{\widetilde{\textsc{cnot}}} & \qw
}
\end{array}
\end{equation}
and since $GS(0) = \textsc{\textsc{swap}}$ a $\widetilde{\textsc{\textsc{swap}}}$ is obtainable by means of the planar circuit
\begin{equation}
\begin{array}{cc}
\Qcircuit @R=1em @C=0.75em {
 & \qw                  & \multigate{1}{GS(0)} & \qw                  & \qw \\
 & \multigate{1}{GS(0)} & \ghost{GS(0)}        & \multigate{1}{GS(0)} & \qw \\
 & \ghost{GS(0)}        & \multigate{1}{GS(0)} & \ghost{GS(0)}        & \qw \\
 & \qw                  & \ghost{GS(0)}        & \qw                  & \qw \\ 
}
\end{array}
=
\begin{array}{cc}
\Qcircuit @R=1em @C=0.75em {
 & \multigate{3}{\widetilde{\textsc{swap}}} & \qw \\
 & \ghost{\widetilde{\textsc{swap}}} & \qw \\
 & \ghost{\widetilde{\textsc{swap}}} & \qw \\
 & \ghost{\widetilde{\textsc{swap}}} & \qw \\
}
\end{array} .
\end{equation}
Since $\textsc{cnot}$, \textsc{\textsc{swap}}, and the above single qubit gates are universal for quantum computation, any circuit from a family recognizing a language in BQP can be represented as a planar $GS$ gate circuit on twice as many qubits and with an at most polynomially larger depth.

\subsection{Gate decompositions}\label{app:gates}
To implement the $GS(\theta)$ gate on superconducting hardware requires us to decompose it into native gates.
In our case this is arbitrary single-qubit rotations and two-qubit number-conserving excitation gates~\cite{Google2020Observation}.
To minimize calibration overhead, we limit ourselves to a fixed two-qubit gate; in all experiments performed this was a controlled-Z gate.
We can write an arbitrary single-qubit rotation in phased-XZ form~\cite{Google2019Quantum}
\begin{equation}
    R(\alpha_x,\alpha_a,\alpha_z) = e^{i(\alpha_z+\alpha_a) Z}e^{i\alpha_x X}e^{-i\alpha_a Z}
\end{equation}
A $GS(\theta)$ gate can be executed at arbitrary $\theta$ on qubits $i$ and $j$ using a combination of $3$ CZ gates and single-qubit rotations
\begin{align}
    GS_{i,j}(\theta)=
    &R_i(\tfrac{\pi}{4},\tfrac{\pi}{4},0)\cdot CZ_{i,j}\nonumber \\
    &\cdot R_i(\tfrac{\pi}{4},\tfrac{-\pi}{4},0)\cdot R_j(\tfrac{\pi}{4}+\tfrac{\theta}{2},\tfrac{-\pi}{4},\tfrac{\pi}{2})\cdot CZ_{i,j}\nonumber\\
    &\cdot R_i(\tfrac{\pi}{4},\tfrac{\pi}{4},0)\cdot R_j(\tfrac{\pi}{4}+\tfrac{\theta}{2},\tfrac{-\pi}{4},\tfrac{\pi}{2})\cdot CZ_{i,j}\nonumber\\
    &\cdot R_i(\tfrac{\pi}{4},\tfrac{-\pi}{4},0).
\end{align}
Alternatively, if one defines a $\sqrt{\textsc{swap}}$ gate to be
\begin{align}
    \sqrt{\textsc{swap}} = \left(\begin{array}{cccc}
    1 & 0 & 0 & 0 \\
    0 & \tfrac{e^{-i\tfrac{\pi}{4}}}{\sqrt{2}} & \tfrac{e^{i\tfrac{\pi}{4}}}{\sqrt{2}} & 0 \\
    0 & \tfrac{e^{i\tfrac{\pi}{4}}}{\sqrt{2}} & \tfrac{e^{-i\tfrac{\pi}{4}}}{\sqrt{2}} & 0 \\
    0 & 0 & 0 & 1
    \end{array}\right),
\end{align}
we can decompose a $GS(\theta)$ gate into two $\sqrt{\textsc{swap}}$ gates and single-qubit Z rotations
\begin{equation}
    GS_{i,j}(\theta) = \sqrt{\textsc{swap}}_{i,j}e^{i\frac{\theta}{2}Z_i}e^{-i\frac{\theta}{2}Z_j}\sqrt{\textsc{swap}}_{i,j}.
\end{equation}
This is similar to the decomposition of a Givens rotation gate into two $\sqrt{I\textsc{swap}}$ and z-rotation gates~\cite{Google2020Hartree}: the $\sqrt{\textsc{swap}}$ and $\sqrt{I\textsc{swap}}$ gates differ by a $e^{i\tfrac{\pi}{4}ZZ}$ rotation, which doubles to yield the $ZZ$ term which separates the Givens and GS rotation (up to a redefinition of angle and single-qubit $Z$ rotations).

It is further possible to decompose a $GS(\theta)$ gate into three $\sqrt{I\textsc{swap}}$ gates and arbitrary single-qubit rotations, though the calculation is more involved.
We start from a decomposition of a \textsc{swap} ($=GS(0)$) gate into three $\sqrt{I\textsc{swap}}$ gates
\begin{widetext}
\begin{align}
    \mathrm{\textsc{swap}}&=\left(e^{-i\frac{\pi}{4}(YI+IY)}\sqrt{\mathrm{i\textsc{swap}}}e^{i\frac{\pi}{4}(YI+IY)}\right)\sqrt{\mathrm{i\textsc{swap}}}\left(e^{i\frac{\pi}{4}(XI+IX)}\sqrt{\mathrm{i\textsc{swap}}}e^{-i\frac{\pi}{4}(XI+IX)}\right)\\
    &=\Big(\exp\left[i\tfrac{\pi}{8}(YY+ZZ)\right]\Big)\exp\left[i\tfrac{\pi}{8}(XX+YY)\right]\Big(\exp\left[i\tfrac{\pi}{8}(XX+ZZ)\right]\Big).
\end{align}
If we perform a basis rotation on the bracketed terms by $\exp(i\alpha Z)$ on the appropriate qubit, we can generate a gate of the form
\begin{equation}
    G(\alpha)=\exp\left[i\tfrac{\pi}{8}(\cos(\alpha)YY + \sin(\alpha)XY)\right]\exp\left[i\tfrac{\pi}{8}(XX+YY+2ZZ)\right]\exp\left[i\tfrac{\pi}{8}(\cos(\alpha)XX-\sin(\alpha)XY\right].
\end{equation}
This function can be expanded to give
\begin{equation}
    G(\alpha)=g_1(\alpha)II+g_2(\alpha)(XX+YY)+g_3(\alpha)(IZ-ZI)+g_4(\alpha)ZZ,
\end{equation}
with $g_i(\alpha)$ the following complex functions
\begin{align}
    g_1(\alpha)&=\frac{1}{\sqrt{2}}\left\{\cos^2(\tfrac{\pi}{8})(\cos^2(\tfrac{\pi}{8})+i\sin^2(\tfrac{\pi}{8}))+i\cos(\alpha)\cos(\tfrac{\pi}{8})\sin(\tfrac{\pi}{8})\frac{(1+i)}{\sqrt{2}}\right.\nonumber\\&+\left.\cos^2(\alpha)\sin^2(\tfrac{\pi}{8})(i\cos^2(\tfrac{\pi}{8})+\sin^2(\tfrac{\pi}{8}))+\sin^2(\alpha)\sin^2(\tfrac{\pi}{8})(\cos^2(\tfrac{\pi}{8})+i\sin^2(\tfrac{\pi}{8}))\right\}\\
    g_2(\alpha)&=\frac{1+i}{4\sqrt{2}}(1+\cos(\alpha))\\
    g_3(\alpha)&=\frac{1}{\sqrt{2}}\left\{\sin(\alpha)\sin(\tfrac{\pi}{8})\cos(\tfrac{\pi}{8})\frac{(1+i)}{\sqrt{2}}+i\sin(\alpha)\cos(\alpha)\sin^2(\tfrac{\pi}{8})(\cos^2(\tfrac{\pi}{8})+i\sin^2(\tfrac{\pi}{8}))\right.\nonumber\\&\left.-i\cos(\alpha)\sin(\alpha)\sin^2(\tfrac{\pi}{8})(i\cos^2(\tfrac{\pi}{8})+\sin^2(\tfrac{\pi}{8}))\right\}\\
    g_4(\alpha)&=\frac{1}{\sqrt{2}}\left\{\cos^2(\tfrac{\pi}{8})(i\cos^2(\tfrac{\pi}{8})+\sin^2(\tfrac{\pi}{8}))-i\cos(\alpha)\cos(\tfrac{\pi}{8})\sin(\tfrac{\pi}{8})\frac{(1+i)}{\sqrt{2}}\right.\nonumber\\
    &\left.+\cos^2(\alpha)\sin^2(\tfrac{\pi}{8})(\cos^2(\tfrac{\pi}{8})+i\sin^2(\tfrac{\pi}{8}))+\sin^2(\alpha)\sin^2(\tfrac{\pi}{8})(i\cos^2(\tfrac{\pi}{8})+\sin^2(\tfrac{\pi}{8}))\right\}.
\end{align}
\end{widetext}
This is of the correct form for our desired gate up to single-qubit $Z$ rotations as long as the phase on $\langle 00|G(\alpha)|00\rangle$, $\langle 11|G(\alpha)|11\rangle$, $\langle 01|G(\alpha)|10\rangle$ and $\langle 10|G(\alpha)|01\rangle$ are equal. One can confirm that all have a phase of $e^{i\tfrac{\pi}{4}}$ for any angle of $\alpha$.
There remains a residual phase on the two on-diagonal elements of $G(\alpha)$,
\begin{align}
    \langle 01|G(\alpha)|01\rangle &= g_1(\alpha) - g_4(\alpha) + 2g_3(\alpha) \nonumber \\ &= A(\alpha) e^{i\phi(\alpha)}e^{i\pi/4}\\
    \langle 10|G(\alpha)|10\rangle &= g_1(\alpha) - g_4(\alpha) - 2g_3(\alpha) \nonumber \\ &= -A(\alpha) e^{-i\phi(\alpha)}e^{i\pi/4},
\end{align}
which can be removed by shifting
\begin{equation}
    G(\alpha)\rightarrow e^{i\phi/4 (IZ-ZI)}G(\alpha)e^{i\phi/4 (IZ-ZI)}.
\end{equation}
The precise value of $\phi$ here is
\begin{align}
    &\phi=-\arctan\Bigg\{\nonumber\\
    &\frac{\sqrt{2}\left(\cos^2(\tfrac{\pi}{8})-\sin^2(\tfrac{\pi}{8})\cos(2\alpha)-\frac{1}{\sqrt{2}}\cos(\alpha)\right)}{\sin(\alpha)+\sin(2\alpha)\sin^2(\tfrac{\pi}{8})}\Bigg\}.
\end{align}
We notice that our formula for $g_2(\alpha)$ only takes positive values. To get the full range of $GS(\theta)$, one can finally send $G(\alpha)\rightarrow e^{i\pi/2 ZI}e^{i\pi/2 IZ}G(\alpha)e^{i\pi/2 ZI}e^{i\pi/2 IZ}$ (again at no extra cost), and solve for $\sin(\theta)=2 g_2(\alpha)$.

\subsection{Scheduling details}\label{app:scheduling}

A key advance in this work was the development of a mapping of our UpCCD ansatz to a 2D grid with local connectivity, such that a) the entire ansatz could be implemented in depth $N/2$ GS gates, and b) all $X_iX_j+Y_iY_j$ terms could be estimated using only $N$ unique mappings.
In this section, we explain this mapping in more detail and prove both a) and b) true.

\begin{figure}
    \centering
    \includegraphics[width=\columnwidth]{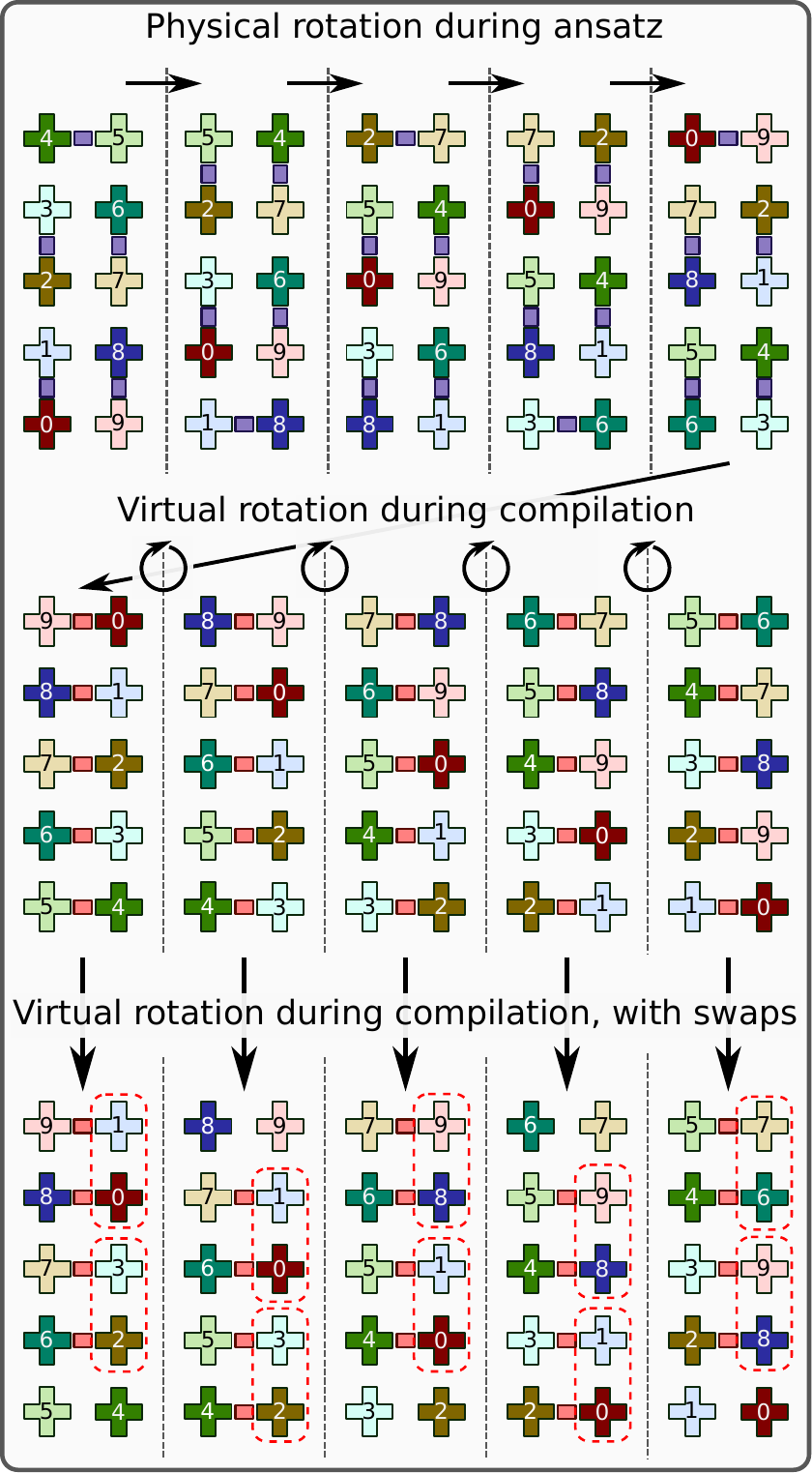}
    \caption{Detail of the scheduling of pair excitation interactions during the UpCCD ansatz, and measurement scheduling. (top) The physical rotations during the execution of \textsc{swap} layers of the UpCCD ansatz, showing that all occupied (even index) and all unoccupied (odd index) orbitals are coupled (purple squares) at some point during the ansatz. (middle) After executing the UpCCD ansatz we have not used the cross-couplers (red squares), allowing us to virtually rotate the entire grid during compilation for the purposes of measurement. This virtual rotation allows all occupied and unoccupied orbitals to be coupled at some step, without any increase in circuit depth. (bottom) The virtual rotation cannot however, bring two occupied or two unoccupied orbitals to nearest-neighbour qubits so that they may be coupled. To achieve this, we require an extra round of \textsc{swap}s (red dashed boxes). One can confirm that across the middle and bottom layers, all pairs of qubits are coupled in at least one configuration.}
    \label{fig:scheduling_detailed}
\end{figure}

Let us first expand the discussion of the implementation of the UpCCD ansatz.
The standard UpCCD ansatz takes the form
\begin{equation}
    U(\theta)=\exp\bigg\{\sum_{p\in \mathrm{unocc}, q\in\mathrm{occ}}\theta_{pq}a^{\dag}_{p\alpha}a^{\dag}_{p\beta}a_{q\alpha}a_{q\beta}-\mathrm{h.c.}\bigg\},
\end{equation}
which when mapped to qubits in the S$_{0}$ approximation, becomes
\begin{equation}
    U(\theta)=\exp\bigg\{\sum_{p\in \mathrm{unocc}, q\in\mathrm{occ}}\theta_{pq}X_pY_q-\mathrm{h.c.}\bigg\}.\label{eq:UpCCD_qubit}
\end{equation}
This in turn can be Trotterized to a product of coherent pair excitations $\exp(\theta_{pq}X_pY_q-\mathrm{h.c.})$.
As mentioned in the main text, the effect of a single $GS$ gate (Eq.~\eqref{eq:GSGate}) in the fermionic picture is to implement a single coherent pair excitation between the spatial orbitals assigned to qubits $i$ and $j$, and then to \textsc{swap} the orbitals.
This means a given orbital is not assigned to a fixed qubit throughout the experiment.
For our implementation of the UpCCD ansatz (see e.g. Fig.~\ref{fig:overview}, bottom) GS gates are applied in layers: between qubits $i,j=2n, 2n+1$ during an even-numbered layer, and between qubits $i,j=2n+1, (2n+2)\% N$ during an odd-numbered layer (for $n=0,\ldots,N/2-1$).
We claim that, at half-filling, any initial assignment of occupied orbitals to odd-numbered qubits and unoccupied orbitals to even-numbered qubits will cause $N/2$ such layers to implement a Trotterized form of Eq.~\eqref{eq:UpCCD_qubit}.
To see this, note that during an even-numbered layer, orbitals sitting on even-numbered (odd-numbered) qubits shift to the right (left) around the ring of qubits, and vice-versa during an odd-numbered layer.
This in turn implies that the empty orbitals, that are initially assigned to an even-numbered qubit, will only ever move to the right around this ring as the ansatz proceeds, as they will be assigned to an odd-numbered qubit on odd-numbered rounds.
Likewise, the filled orbitals will only ever move to the left around this ring.
As an occupied orbital must cross (and thus interact with) every unoccupied orbital before it encounters the same one twice, this implies that the first $N/2$ layers of our UpCCD ansatz will yield precisely the $(N/2)^2$ coherent pair excitations between each unoccupied and each occupied orbital, as required.
(This is demonstrated for $10$ qubits in Fig.~\ref{fig:scheduling_detailed}[top].)

Let us now consider the measurement of non-local $X_iX_j+Y_iY_j$ terms as performed in this work.
As mentioned in the main text, these are diagonalized on a pair of orbitals $i,j$ via a $GS(\pi/4)$ rotation, which necessitates the orbitals be on neighbouring qubits.
The $GS(\pi/4)$ rotation maps the operators $Z_i, Z_j\rightarrow D_{ij}^+, D_{ij}^-$, where we define
\begin{equation}
    D_{ij}^{\pm} = \frac{1}{2}\Big[Z_i+Z_j\pm \big(X_iX_j+Y_iY_j\big)\Big]\label{eq:Dij}
\end{equation}
As we implement our ansatz on a $2\times N/2$ grid, qubit $i$ is not only connected to qubit $(i\pm 1)\% N$; the cross-links in the grid connect qubit $i$ and qubit $N-i-1$.
Moreover, note that our ansatz remains unchanged if we perform the cyclic permutation $i\rightarrow (i+k) \% N$; that is, we shift our orbital assignments, and all ansatz gates and parameters, around the ring of qubits.
(Note that this technically \textsc{swap}s the definition of even and odd layers when $k$ is odd: after this permutation the first layer of GS gates will be between qubits $i,j=2n+1, (2n+2)\% N$.)
Following this permutation, cross-links will connect the orbital that was on qubit $i$ to that which was on qubit $[N-(i+2k)-1]\% N$.
One can confirm that by running over $k=0,\ldots,N/2-1$ we find $N/2$ cyclic permutations, such that each occupied orbital is coupled to each unoccupied orbital by a cross-link for exactly one permutation.
With just the cyclic permutation operation and no additional \textsc{swap} gates, this gives $N/2$ unique circuits that allow for measurement of all $D_{ij}^{\pm}$ where $i$ is occupied and $j$ unoccupied.
To obtain the circuits that couple occupied orbitals to occupied orbitals (and unoccupied to unoccupied), we require an additional layer of \textsc{swap} gates.
(This is unavoidable given our initial ansatz ordering: all occupied qubits are connected by an even number of couplings, so a further direct coupling would yield an odd-order cycle, which does not exist on a square lattice.)
After each permutation $k$ above, we perform \textsc{swap}s between qubits $l + (k\% 2), l + (k\% 2) + 1$ for $0\leq l < N/4$.
One can confirm that this yields $N/2$ circuits such that a coupling between any pair of occupied orbitals can be achieved in one such circuit.
(In this second set of circuits, some qubits are not coupled, and were not used to estimate expectation values in this experiment.)
This can be confirmed by a visual inspection of Fig.~\ref{fig:scheduling_detailed}[bottom].
Code that implements this scheduling has also been uploaded to ReCirq~\cite{ReCirq}.

\subsection{Scheduling of EV circuits}\label{app:ev-scheduling}
In this section we outline the additional experimental details required to implement EV in this work.
We implemented control-free EV for $O=Z_i,$ or $Z_iZ_j$, or $D_{ij}^+$(Eq.~\eqref{eq:Dij}), using a vacuum reference state $|\phi\rangle=|00\ldots\rangle$.
We prepare the superposition $\frac{1}{\sqrt{2}}(|\psi\rangle+|\phi\rangle)$ by acting the UpCCD ansatz circuit (Eq.~\eqref{eq:UpCCD}) on the cat state $\frac{1}{\sqrt{2}}(|0000\ldots\rangle+|0101\ldots\rangle)$ (see Fig.~\ref{fig:overview}[middle, 'Echo verification'] and Fig.~\ref{fig:overview}[bottom]).
Then, all operators $O$ were implemented by compiling them to a virtual $Z$ rotation on a single qubit.
Finally, to measure $\langle\Psi|\phi\rangle\langle\psi|\Psi\rangle$, we inverted the UpCCD circuit and cat state preparation.
This maps the desired matrix element $|\phi\rangle\langle\psi|\rightarrow |0\rangle\langle 1|\otimes |00\ldots\rangle\langle 00\ldots|$, allowing its measurement via single-qubit rotation on a single 'measurement' qubit, and readout of all qubits in the computational basis.
(Reading out all qubits is essential, as we record an estimate of $0$ for the measurement of $\langle\Psi|\phi\rangle\langle\psi|\Psi\rangle$ unless all qubits other than the measurement qubit read out $0$.)

Our implementation of $O$ as a virtual $Z$ rotation allows us to replace $O\rightarrow O^{\alpha}=\cos(\alpha)+i\sin(\alpha)O$ to remove our susceptibility to a uniform background magnetic field $e^{ih\sum_jZ_j}$.
Such a field transforms Eq.~\eqref{eq:CFEV_expval} to $\tfrac{1}{2}\langle\psi|O^{\alpha}|\psi\rangle e^{i(\phi + hN/2)}$; fitting this to three points of $\alpha$ allows us to simultaneously estimate $h$, the fidelity $F$, and $\langle O\rangle$.
(This is preferable to independent calibration as $h$ fluctuates with a $1/f$ spectrum.)

Some further experimental optimization was made for the EV circuit that was not available for the VQE or VD circuits.
Many of the gates in the final EV circuit [Fig.~\ref{fig:overview}, bottom] cancel to the identity, as the second half of the circuit is the inversion of the first half.
We identify and prune these to increase the overall circuit fidelity, and insert echo pulses into the resulting empty space.
We further compile an echo pulse for the entire second half of the EV circuit [this can be done as $XX\cdot GS(\theta)\cdot XX = GS(\theta)$].
To unbias readout, we measure the single measurement qubit in the $\pm X$ and $\pm Y$ bases.
This, combined with the additional circuits to remove a background magnetic field, raises the total number of circuits to $12N^2$ ($6N^2+6N$) to estimate the expectation value of our chemistry (RG) Hamiltonian.
Shots were distributed across these circuits following the term weight in the Hamiltonian with some additional restrictions imposed by classical readout hardware (see App.~\ref{app:shot_distribution_expt}).

\section{Data processing}

\subsection{Optimal linear combinations of non-independent expectation values}\label{app:optimal-linear-combination}
Once we have used our variational ansatz to prepare an approximation $\rho(\boldsymbol{\theta})\sim|\psi(\boldsymbol{\theta})\rangle\langle\psi(\boldsymbol{\theta})|$ to the ground state of our target problem, it remains to measure the quantum device to estimate the energy (or other properties of the state).
As our devices are heavily coherence limited, rather than attempting to perform this estimation in a single shot, we write our Hamitonian as a sum of simpler terms
\begin{equation}
    H=\sum_i c_iQ_i,\label{eq:ham_decomp}
\end{equation}
and estimate the expectation value of each such term
\begin{equation}
    E=\mathrm{Trace}[H\rho]=\sum_ic_i\mathrm{Trace}[Q_i\rho].\label{eq:energy_estimate}
\end{equation}
(Note that we are not placing any restrictions on $Q_i$ at this point.)
The method in which we estimate $\mathrm{Trace}[Q_i\rho]$ will depend on which error mitigation methods are being implemented.
However, all schemes will return a set of estimates of $\mathrm{Trace}[Q_i\rho]$ with a covariance matrix 
\begin{equation}
\Sigma_{i,j}=\mathrm{Covar}\Big[\mathrm{Trace}[Q_i\rho],\mathrm{Trace}[Q_j\rho]\Big].
\end{equation}
In this experiment, it turns out that our choice of $\{Q_i\}$ will not be linearly independent.
The reason for this is post-selection: we desire our choice of $\{Q_i\}$ to allow us to measure $S_z=\sum_iZ_i$ for each experiment.
To achieve this, we measure the operators $Z_j, Z_jZ_k,$ and $D_{jk}^{\pm}$ (Eq.~\eqref{eq:Dij}), but $D_{jk}^{+}+D_{jk}^{-}=Z_j+Z_k$.
This leaves us with a degree of freedom in our choice of $c_i$ that we may optimize upon, once a dataset is taken.

In order to choose $c_i$, we perform a constrained Lagrangian minimization.
Our target cost function is the variance on the estimate in Eq.~\eqref{eq:energy_estimate}
\begin{equation}
    \mathrm{Var}(E)=\sum_{i,j}c_i\Sigma_{i,j}c_j,
\end{equation}
subject to Eq.~\eqref{eq:ham_decomp} as a constraint.
Let us fix some linearly independent basis of operators $\{P_j\}$ (e.g. Pauli operators), and we can write $H=\sum_jh_jP_j$, and $Q_i=\sum_jq_{i,j}P_j$.
(As $P_j$ is a basis, we have no freedom in our choice of $h_j$ or $q_{i,j}$.)
Our constraints then take the form
\begin{equation}
    \sum_{i}c_iq_{i,j}=h_j.
\end{equation}
This can be written as a Lagrange multiplier, yielding a Lagrangian
\begin{equation}
    \mathcal{L}=\sum_{i,j}c_i\Sigma_{i,j}c_j-\sum_{j}\Big(\sum_ic_iq_{i,j}-h_j\Big)\lambda_j.
\end{equation}
Differentiating with respect to $c_i$ and setting equal to $0$ yields (using the fact that $\Sigma_{i,j}$ is a positive matrix)
\begin{equation}
    2\sum_j\Sigma_{i,j}c_j-\sum_jq_{i,j}\lambda_j=0\rightarrow c_j.
\end{equation}
Recognising this as a vector equation, as the matrix $\Sigma$ is invertible we have
\begin{equation}
    \mathbf{c}=\frac{1}{2}\Sigma^{-1}q\boldsymbol{\lambda}.
\end{equation}
Here, $\boldsymbol{\lambda}, \mathbf{c}$ are vectors containing the $\lambda_j$ and $c_j$ components, and $q$ and $\Sigma$ are matrices containing the $q_{i,j}$ and $\Sigma_{i,j}$ components respectively.
Substituting this into our Lagrangian yields
\begin{equation}
    \mathcal{L}=-\frac{1}{4}\boldsymbol{\lambda}^Tq^T\Sigma^{-1}q\boldsymbol{\lambda}+\boldsymbol{\lambda}^T\mathbf{h}.
\end{equation}
Then, differentiating with respect to $\lambda$ (and using the fact that $q^T\Sigma^{-1}q$ is Hermitian), we have
\begin{align}
    &q^T\Sigma^{-1}q\boldsymbol{\lambda}=2\mathbf{h}\rightarrow \boldsymbol{\lambda}=2(q^T\Sigma^{-1}q)^*\mathbf{h}\\&\rightarrow\mathbf{c}=\Sigma^{-1}q(q^T\Sigma^{-1}q)^*\mathbf{h},
\end{align}
where here, $*$ denotes the Moore-Penrose pseudoinverse of a matrix.

In practice, though we find that this produces low-variance estimates, it is unstable to uncertainty in our estimate of $\Sigma$.
As such, we set $\Sigma=I$ for the purposes of determining $\mathbf{c}$ (which corresponds to assuming that all uncertainties are equal and all covariances are $0$).
This yields
\begin{equation}
    \mathbf{c}=q(q^Tq)^*\mathbf{h}.
\end{equation}

\subsection{Error propagation and bootstrapping}\label{app:errorbars}
The exact form of the error for raw and post-selected VQE is well-known.
The covariance between estimates of the expectation value of two reflection operators $P_i$ and $P_j$, estimated simultaneously from $M$ repeated preparations and measurements on a target state, is given by
\begin{equation}\label{eq:reflection_covariance}
    \mathrm{Covar}[\langle P_i\rangle\langle P_j\rangle] = \frac{\langle P_iP_j\rangle - \langle P_i\rangle\langle P_j\rangle}{M}.
\end{equation}
The resulting number can be substituted into the propagation of variance formula described in App.~\ref{app:optimal-linear-combination} above to get an estimate of the variance in energy, whilst errors in order parameters can be obtained by propagation of variance through ($\Delta = \frac{1}{N}\sum_{j,\sigma}\sqrt{\langle n_{j\sigma}^{2}\rangle-\langle n_{j\sigma}\rangle^2}$)
\begin{align}
    \frac{\partial \Delta}{\partial \langle n_{j\sigma}\rangle}&=\frac{1}{2N}\frac{1-2\langle n_{j\sigma}\rangle}{\sqrt{\langle n_{j\sigma}\rangle -\langle n_{j\sigma}\rangle^2}}\\
    \mathrm{Var}[\Delta] &= \sum_{j\sigma}\frac{1}{2N}\frac{(1-2\langle n_{j\sigma}\rangle)^2}{\langle n_{j\sigma}\rangle - \langle n_{j\sigma}\rangle^2}\mathrm{Var}[\langle n_{j\sigma}\rangle].
\end{align}
We note that this diverges when $\langle n_{j,\sigma}\rangle\rightarrow 0,1$, which is what happens at $g=0$ when $\Delta\rightarrow 0$.
We suggest that this explains the peak in the observed experimental error in the order parameter around $g=0$ somewhat.

The experimental covariances (Eq.~\eqref{eq:reflection_covariance} for raw and postselected VQE when $i\neq j$ were corrupted during data taking; these terms were set to $0$ when generating error bars for Fig.~\ref{fig:cyclobutene_fig} and Fig.~\ref{fig:RG_figure}.
As said error bars are negligible due to the large number of samples used in this experiment, more complex recovery procedures will not noticeably change the figures.

For echo verification and virtual distillation, the formulas for variance are more complicated (the form derived in Ref.~\cite{Polla2022Optimizing} is not appropriate here due to our fitting to remove a background magnetic field).
Instead, error bars were determined by bootstrapping the raw data (resampling with replacement and taking the sample standard deviation) over $100$ and $25$ samples respectively.
This was made complicated due to data loss during the the EV experiment.
Arrays were stored of the verified expectation value of $X$ and $Y$ on the measurement qubit; this is insufficient to recover the shot distribution, as the EV measurement can return three values; $+1$, $-1$, and $0$~\cite{Obrien2021Error}.
Counts $M_{+}$, $M_{-}$, and $M_0$ of these values were approximated from these expectation valus using the estimated EV fidelity $F_{\mathrm{EV}}$ and the fact that in the absence of error
\begin{equation}
\frac{M_0}{M} = 1 - |\langle U\rangle|^2,
\end{equation}
where $U$ is the operator to be estimated via EV, and $M=M_++M+-+M_0$.
Assuming the fraction $1-F_{\mathrm{EV}}$ fails verification, we have $M_0=M(1-F_{\mathrm{EV}}|\langle U\rangle|^2)$, and $M_+$ and $M_-$ may be then distributed so that $(M_+-M_-)/M$ is the observed expectation value on the measurement qubit.
(This can result in $M_+<0$ or $M_-<0$, in which case we reduced $M_0\rightarrow M_0 - 2\min(M_+,M_-)$, $M_+,M_- \rightarrow M_+ +\min(M_+,M_-), M_-+\min(M_+,M_-)$.) 

\section{Quantum run time estimation}\label{app:extrapolation}
In this section, we estimate the cost of running our quantum experiments in terms of number of experiments (shots) and the wall-clock time.
This is done in terms of established theoretical cost estimates~\cite{Wecker2015Progress,Rubin2018Application}, which allows us to compare between this and what was implemented in the actual experiment in Fig.~\ref{fig:scaling_figure} (top).
This is further necessary for the tuning of hyperparameters of the variational optimizers we will present in App.~\ref{app:optimization}.

\subsection{Wall-clock model} \label{app:wall-clock_time_model}
The cost of running a set of experiments on real-world hardware in terms of the actual time spent or \textit{wall-clock time} is not a linear function of the number of samples used.
We can estimate the wall-clock time as a function of three parameters.
First, the number $a$ of calls to the device from the computed executing the Cirq code.
Second, the number $b$ of distinct circuits the device needs to execute.
Third, the total number of shots $c$ spend over all calls and distinct circuits.
We found empirically that for the device used the time for a call to the device from Cirq is about $1s$.
In each call, a batch of distinct circuits can be sent to the device for execution but re-programming the device to execute a different circuit takes $0.042s$.
When estimating energy expectation values or order parameters, all circuits that need to be executed are known in advance and thus can be sent in a single batch.
However, during variational optimization, depending on the optimizer used, at least one call to the device per epoch is needed, because the quantum circuits to be executed next depend on the step taken by the optimizer, which in turn depends on the measurement results of the first batch.
This needs to be taken into account when comparing the performance of different optimizers.
The time per shot was found to be $1*10^{-5}$, essentially independently of the circuit depth (probably because it is dominated by readout, reset, and time needed by the control electronics).
The number of shots $c$ of a circuit can currently only be set for a batch as a whole, which in practice limits our ability to distribute shots in a way that would minimize the variance of the resulting estimator.
The total wall-clock time this takes the form
\begin{equation}
 t_{\mathrm{wall}} = a * 1s + b * 0.042s + c * 5*10^{-5}s\label{eq:wall_clock}
\end{equation}

\subsection{Hamiltonian decomposition schemes}\label{app:measurement_methods}
In this section we define the different options chosen to decompose a Hamiltonian into terms $Q_i$ for measurement.
We are free to group the measurement of all $Q_i$ terms together, as long as they commute and an appropriate diagonalization circuit is found.
For our numerics, we study the following measurement strategies:
\begin{itemize}
    \item \emph{Termwise} --- we take individual Pauli operators $Q_i\in\{Z_j,X_iX_j,Y_iY_j,Z_iZ_j\}$, and measure each $Q_i$ using an independent measurement circuit.
    \item $XX+YY$ --- we first measure all $Q_i\in\{Z_j,Z_iZ_j\}$ in a single-shot measurement. Next, we measure $Q_i\in\{X_jX_k+Y_jY_k\}$, grouping disjoint pairs $j,k$ together into $N$ sets following the scheduling outlined in Sec.~\ref{app:scheduling}. This mostly matches the scheme used for the estimation of expectation values Raw VQE, PS-VQE and PS-VD in Fig.~\ref{fig:RG_figure}, and the scheme used for the PS-VQE estimates in Fig.~\ref{fig:cyclobutene_fig}.
    \item $XX+YY+IZ+ZI$ --- here, we draw $Q_i\in\{Z_i,Z_iZ_j,D_{ij}^{+}\}$ ($D_{ij}^{\pm}$ is defined in Eq.~\eqref{eq:Dij}), and measure each term separately. This matches the measurement scheme used for EV. As the operators chosen are Hermitian and unitary, the EV variance is equivalent to the standard tomography variance~\cite{Polla2022Optimizing}. By choosing only $D_{ij}^+$ and not adding $D_{ij}^-$ to our set of measurements, the set $\{Q_i\}$ becomes linearly independent, and so the relative coefficients $c_i$ (Eq.\eqref{eq:energy_estimate}) are fixed.
\end{itemize}

Some notable differences occur between the numerical estimates made here and those implemented on hardware.
(Ultimately the number of shots in the experiment was chosen to be low enough that the experimental error mostly dominated, rather than being optimized based on preliminary calculations.)
In the experiment, additional estimates of $Z_j+Z_k$ were extracted alongside each measurement of $X_jX_k+Y_jY_k$ and combined using the techniques outlined in App.~\ref{app:optimal-linear-combination}.
Each circuit was repeated with and without an additional $\pi$ pulse on all qubits to unbias readout noise.
In the experiment, the shot distribution was chosen to be uniform for VQE (40,000 per circuit) and VD (100,000 per circuit).
(One can observe in Fig.~\ref{fig:scaling_figure} that an excess of shots was taken in both cases.)
For EV, shots were distributed according to the relative weight of $Q_i$ in the Hamiltonian.
However, this was made more complicated by a technical requirement that all shots executed in a single batch must have the same number of repetitions.
To accommodate this, the number of shots used to estimate a single $Q_i$ was rounded up or down to be an integer multiple of a fixed base (40000).
Furthermore, as mentioned in App.~\ref{app:ev-scheduling}, when performing EV to estimate each $Q_i$, $12$ unique circuits were run to cancel out a background magnetic field and depolarize readout noise.

\subsection{Shot distribution}\label{app:shot_distribution_expt}
Once the decomposition of the Hamiltonian is decided, the variance of the resulting energy estimator further depends on how the available shot budget is distributed over the $Q_i$ (or the jointly measurable groups of $Q_i$).
In principle, estimates of the (co-)variances from a small number of shots, from previous measurements at close by VQE parameter values according to \eqref{eq:reflection_covariance}, or in adaptive schemes can be used to distribute shots in an asymptotically optimal way to reduce the variance.
In practice, one is limited by the overhead of calling the device and limitations in setting the shots for individual circuits inside a batch (see Section~\ref{app:wall-clock_time_model}).

In our estimates we assume the ability to allocate shots per distinct circuit (and not only on a per-batch basis) but only consider the two non-adaptive shot distribution schemes that need no input from the quantum computer: 

\begin{itemize}
\item \emph{Uniform Distribution} --- distribute the total shot budget per expectation value uniformly over the term groups, i.e., spend the same number of shots on each group of $Q_i$ that are measured jointly according to the decomposition scheme.

\item \emph{According to term group weights} --- the total shot budget is distributed proportional to the coefficients $|c_i|$ in Eq.~\ref{eq:energy_estimate}.
In the case that multiple $Q_i$ are measured, shots are distributed according to $[\sum_i|c_i|^2]^{1/2}$.
This is optimal in case the variances of all $Q_i$ are equal and covariances vanish \cite{Huggins2019Efficient}.
(Note that as the three measurement strategies measure only linearly-independent operators, the $|c_i|$ can be fixed prior to measurement.)
\end{itemize}

\subsection{Wall-clock time extrapolation to large system sizes}
\begin{figure}[tb]
    \centering
    \includegraphics[width=\linewidth]{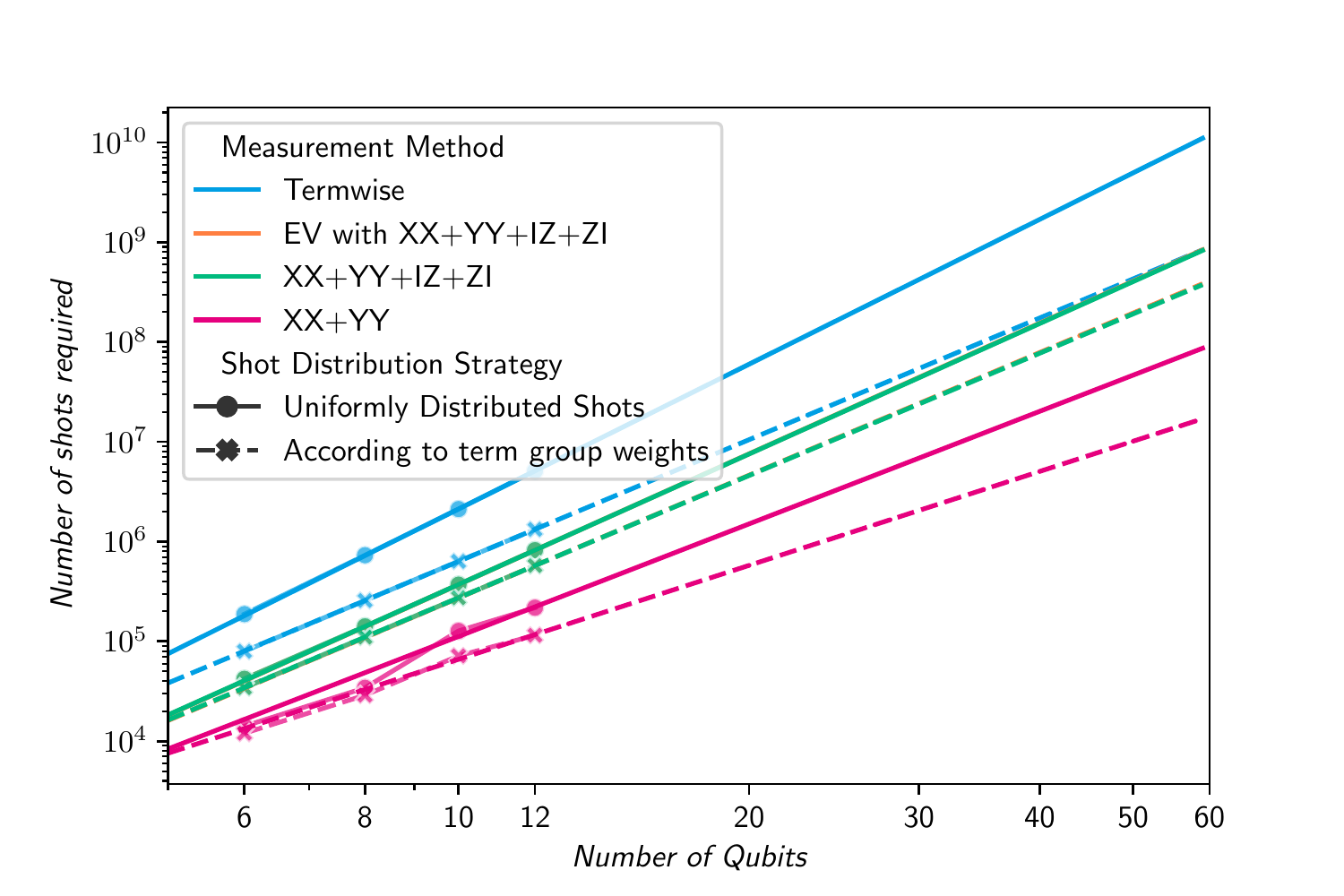}
    \includegraphics[width=\linewidth]{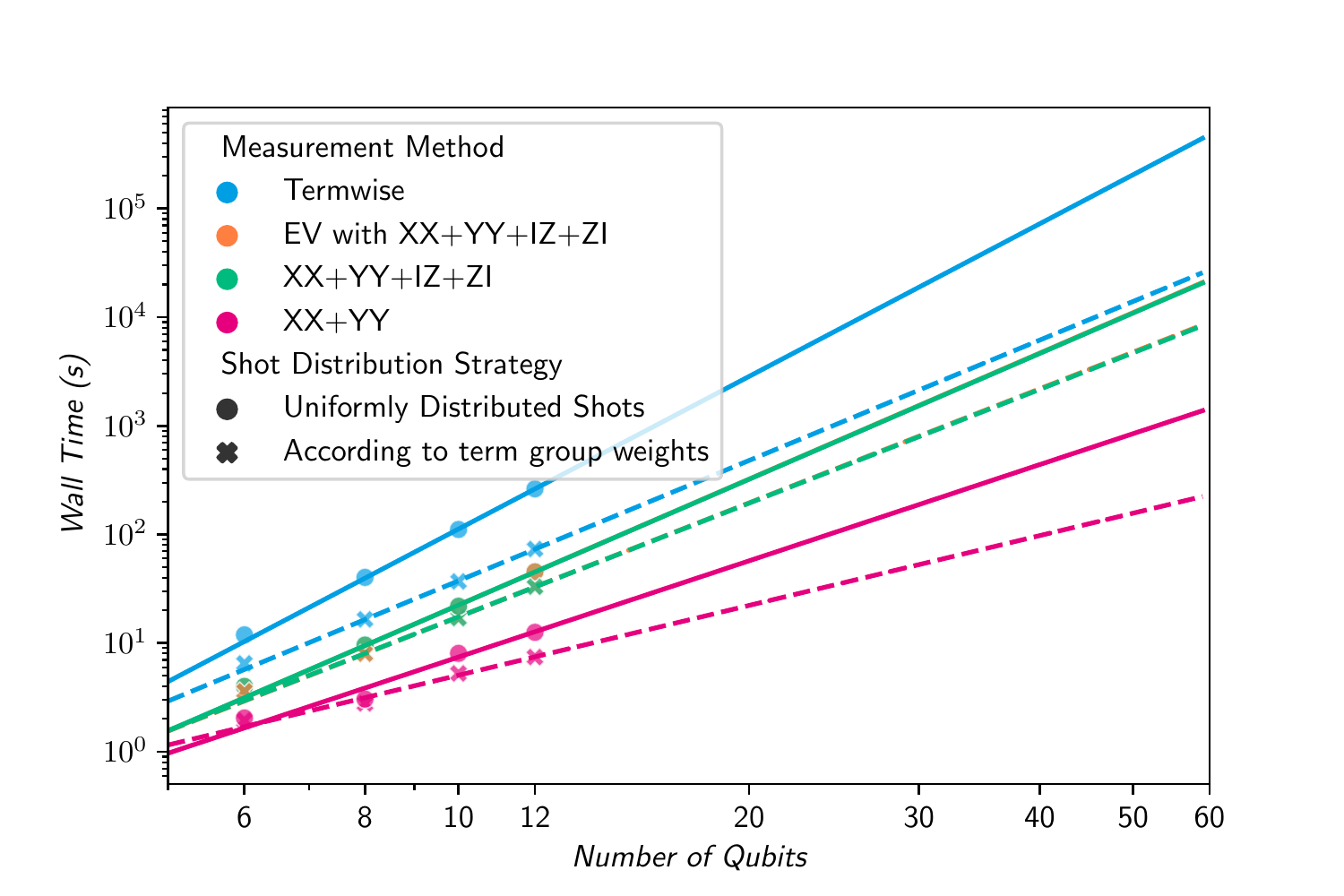}
    \caption{Number of shots (top) and wall-clock time (bottom) required to estimate the ground-state energy of the RG Hamiltonian at $g=-0.9$ to within $0.1\mathrm{a.u.}$.
    Estimates were obtained by standard Lagrangian optimization~\cite{Rubin2018Application} using term distributions defined in App.~\ref{app:shot_distribution_expt} and measurement strategies defined in App.~\ref{app:measurement_methods}. Wall-clock time model is described in App.~\ref{app:wall-clock_time_model}. The orange and green data points for the $XX+YY+IZ+ZI$ scheme with and without EV coincide.}
    \label{fig:number_of_shots_and_wall_clock_extrapolation}
\end{figure}

Using the protocol described in the previous sections, we are able to estimate the cost of measuring the RG model at $g=-0.9$ to within $0.1$ a.u. as we enlarge the system size, while optimizing our shot distribution (App.~\ref{app:shot_distribution_expt}) for the wall-clock time (App.~\ref{app:wall-clock_time_model}).
In Fig.~\ref{fig:number_of_shots_and_wall_clock_extrapolation}, we plot the cost in terms of the number of shots (top) and in wall-clock time for superconducting hardware (bottom).
Estimates are performed for $N=6, 8, 10, 12$; given good fit to a line on a log-log plot and that a powerlaw scaling is expected, we are able to extrapolate this to larger system sizes.
This data was combined with fidelity estimates for a $10$-qubit system (Fig.~\ref{fig:scaling_figure}[bottom-right]) to yield the curves in Fig.~\ref{fig:scaling_figure}[top].

\section{Variational optimization of a 6-qubit experiment}\label{app:optimization}

\subsection{The Conjugate Model Gradient Descent optimizer}
\label{ConjugateMGD}
The large number of evaluations needed for ansatz parameter optimization on quantum hardware is a major impediment towards keeping the cost of variational algorithms (in terms of wall-clock time) manageable.
To mitigate the overhead incurred by sending jobs to and receiving results from a device (see App.~\ref{app:wall-clock_time_model}), it is beneficial if the optimizer can request a batch of cost-function evaluations at once before making a step.
In \cite{Sung2020Models}, surrogate model based optimizers were found to have good performance under this constraint.
Here, a quadratic model function was fitted to present and past expectation value estimates in the vicinity of the current ansatz parameter vector.
Then, after making a step, a batch of circuits corresponding to points in the vicinity of the new parameter vector were evaluated and the stepping procedure repeated.
In this appendix we develop a natural extension of this procedure, by combining it with the conjugate descent algorithm.

The Conjugate Model Gradient Descent optimizer is a surrogate model-based optimization algorithm, with the additional improvement that the gradient which is extrapolated from fitting a quadratic model to the cost function, is used in the framework of conjugate gradient descent to make a step in the parameter space.
The Conjugate Gradient Descent method was developed by Hestenes and Stiefel in~\cite{Hestenes1952Methods}.
For the special case of quadratic cost functions over an $n$ dimensional space, conjugate gradient methods can be proven to converge in $n$ iterations \cite{Daniel1967}.
In practice the conjugate gradient method is found to work well for cost functions that are locally approximately quadratic.
This is an assumption one anyway needs to make when using quadratic model based optimizers and thus makes it natural to combine conjugate gradient descent and quadratic surrogate model based optimizers with quadratic model functions.
In conjugate gradient descent the steps are taken in the direction of the so-called conjugate gradient $s_n = g_n + \beta_n s_{n-1}$, where $g_n$ is the gradient (in our case the one of the quadratic model fitted to the samples around the current position) and $\beta_n$ is a scaling factor.
The formula for the n-th iteration scaling factor is given in \cite{Fletcher1964Function} as:
\begin{math}
\beta^{FR}_n=\frac{g_n^Tg_n}{g_{n-1}^T g_{n-1}}
\end{math}.
With $s_0=g_0$ fixed for the first step of the algorithm.

In Alg.~\ref{alg:ConjModelGrad}, we present our Conjugate Model Gradient Descent algorithm.
We assume in this algorithm access to a (noisy) oracle to the target function $f$ to be optimized.
In practice, this is given by a call to a quantum device with a target error, that must be made small enough to enable convergence.
\begin{algorithm}[H]
\caption{Conjugate Model Gradient Descent}\label{alg:ConjModelGrad}
\floatname{algorithm}{Procedure}
\renewcommand{\algorithmicrequire}{\textbf{Input:}}
\algorithmicrequire{Initial point $x_0$, learning rate $\gamma$, sample radius $\delta$, maximum iterations $n$, number of evaluations per iteration $k$, rate decay exponent $\alpha$, stability constant A, sample radius decay exponent $\xi$, tolerance $\varepsilon$, oracle for function $f$.}
\begin{algorithmic}[1]
\State Initialize lists L, L'
\State Initialize a list G
\State Initialize a list H
\State Let $x$ $\leftarrow$ $x_0$
\For{m in 0\ldots n}
\State Let $\delta'$ $\leftarrow$ $\delta/(m+1)^{\xi}$
\State Sample $k$ points uniformly at random from the $\delta'$-neighborhood of $x$ to generate a set $S$
\For {each $x'$ in $S\cup\{x\}$}
\State Add $(x', f(x'))$ to $L$
\EndFor
\State Clear list $L'$
\For{each tuple $(x',y')$ in $L$}
\If{$|x'-x|<\delta$}
\State Add $(x',y')$ in $L'$
\EndIf
\EndFor
\State Fit $f(x)=x^T A x + bx + c\sim y$ to the points $(x,y)$ in $L'$ using least squares linear regression.
\State Let $g_m$ be the gradient of $f$ at $x$ (i.e. $g_m=b$).
\If{$|g_m|<\varepsilon$}
\State \Return $x$
\EndIf
\If{$m=0$}
\State Let $h_0$ $\leftarrow$ $g_0$
\Else
\State $\beta_{m} \leftarrow g_m^{T} g_m/g_{m-1}^{T} g_{m-1}$
\State $h_{m} \leftarrow g_{m} +\beta_{m}h_{m-1}$
\EndIf
\State $\gamma'=\gamma/(m+1+A)^{\alpha}$
\State Add $g_m$ to the list G
\State Add $cg_m$ to the list H
\State Let $x \leftarrow x -\gamma' \cdot h_m$
\State Let $m \leftarrow m+1$
\EndFor
\State \Return $x$
\end{algorithmic}
\end{algorithm}

\subsection{Hyperparameter tuning}
For all experiments shown in the main text, parameters were obtained by a noiseless simulation using L-BFGS-B, starting from $\boldsymbol{\theta}=0$.
We find that in the absence of experimental noise, gradient-based optimizers such as L-BFGS-B converge well to an optimal solution.
However, this is not the case in the presence of experimental or sampling noise, which can make many estimators unstable.
This has lead previous efforts towards stochastic-based~\cite{Stanisic2021Observing} or model-based~\cite{Google2020Hartree} optimizers, including the Model Gradient Descent optimizer that we have based our conjugated Model Gradient Descent algorithm upon in the previous section.

\begin{figure}[tb]
    \centering
    \includegraphics[width=\linewidth]{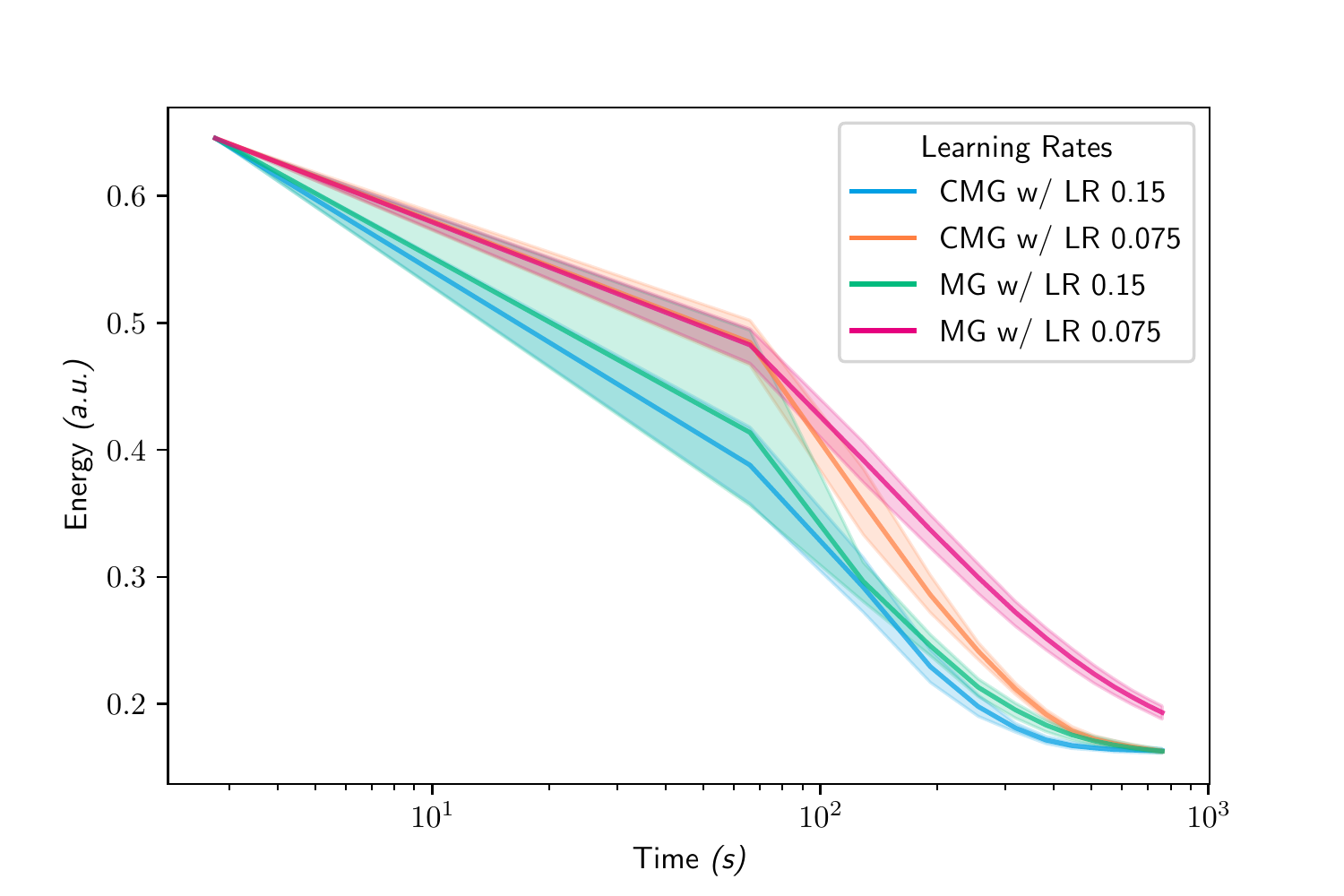}
    \caption{In this figure, the distance between the energy that was calculated in the noiseless regime by the L-BFGS-B optimizer is plotted over the wall-time required to reach such resolution. The system displayed is a RG Hamiltonian for 6 qubits, with a coupling constant of $g=-0.9$, consisting of 10 runs and the lines being the average of those runs for the Conjugate Model Gradient Descent (CMG) and the Model Gradient Descent (MG) found in \cite{Sung2020Models}.}
    \label{fig:nact6_g=-0.9_CMG_MG_convergence}
\end{figure}

\begin{figure}[tb]
    \centering
    \includegraphics[width=\linewidth]{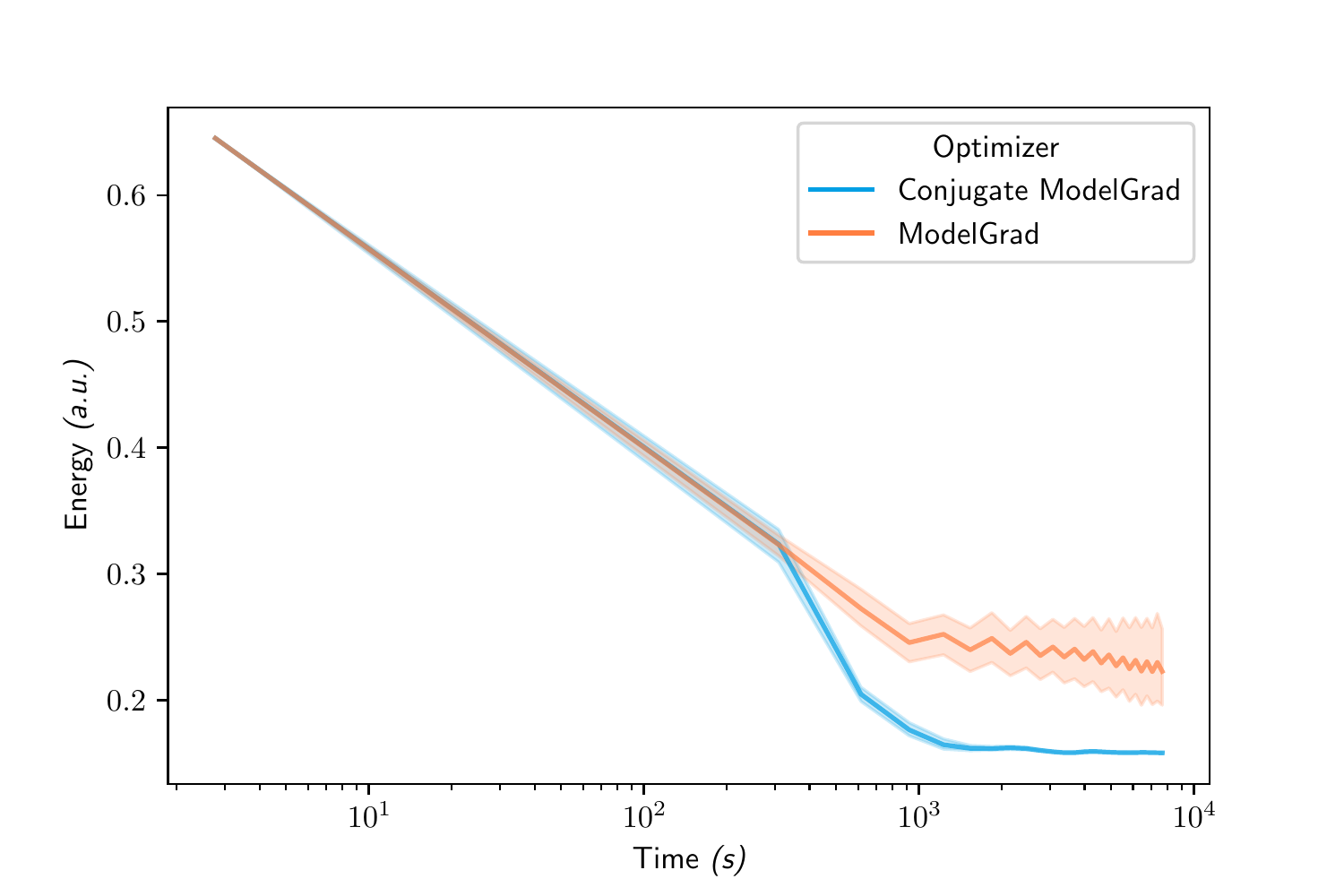}
    \caption{Comparison of the two variants of the Model Gradient Descent optimizer, where in this case the number of samples used to construct the quadratic model one uses is higher, namely $k=4(N+1)(N+2)$, allowing for the performance difference of the two optimizers to become more pronounced, with Conjugate Model Gradient Descent managing to consistently converge at better parameters, while Model Gradient Descent gets stuck in a local minimum, averaged over 10 runs for each optimizer, for 25 maximum iterations.}
    \label{fig:high_sample_number_differences}
\end{figure}

The hyperparameters used in the Conjugate Model Gradient Descent algorithm were either chosen through an intuitive approach due to their physical meaning, such as the sample radius due to the parameters being perturbed by a random variable from [-0.25,0.25] and the maximum number of iterations to stay within reasonable wall-time, or via grid search for the rest of the hyperparameters, ensuring they are good enough to work for the variety of cases examined while avoiding overfitting. Explicitly, the hyperparameters used are included in Table~\ref{tab:hyperparameters_of_cmg}:
\begin{table}[H]
    \centering
    \begin{tabular} {|c c|}
     \hline
     Hyperparameter&  Values Used\\
     \hline\hline
     Sample radius $\delta$ & $1.0$\\
     Learning rate $\gamma$ & $0.15$\\
     Stability constant A & $0$\\
     Sample number $k$ & $0.409(N+1)(N+2)$\\
     Sample radius decay exponent $\xi$ & $0$ \\
     Rate decay exponent $\alpha$ & $0.2$\\
     Maximum evaluations $n$ & $12$ \\
     \hline
\end{tabular}
    \caption{Hyperparameters for the Conjugate Model Gradient Descent optimizer during the experiment, where $N$ is the number of parameters in the optimization. These were also used in the later comparisons with Model Gradient Descent, to gauge performance in equal footing.}
    \label{tab:hyperparameters_of_cmg}
\end{table}
The hyperparameters we chose work equally well for Model Gradient Descent, as well our Conjugate variant.
We found that the conjugate method can sometimes speed up convergence or help prevent getting stuck in local minima.
We illustrate this with two examples:
In Fig~\ref{fig:nact6_g=-0.9_CMG_MG_convergence} how Conjugate Model Gradient Descent has the ability to speed up in case the initial learning rate was chosen to be too small. 
In Fig.~\ref{fig:high_sample_number_differences} we show an example of hyper parameters for which Conjugate Model Gradient Descent is able to converge to a lower minimum than plain Model Gradient Descent.


\subsection{Experimental results}
\begin{figure}
    \centering
    \includegraphics[width=\columnwidth]{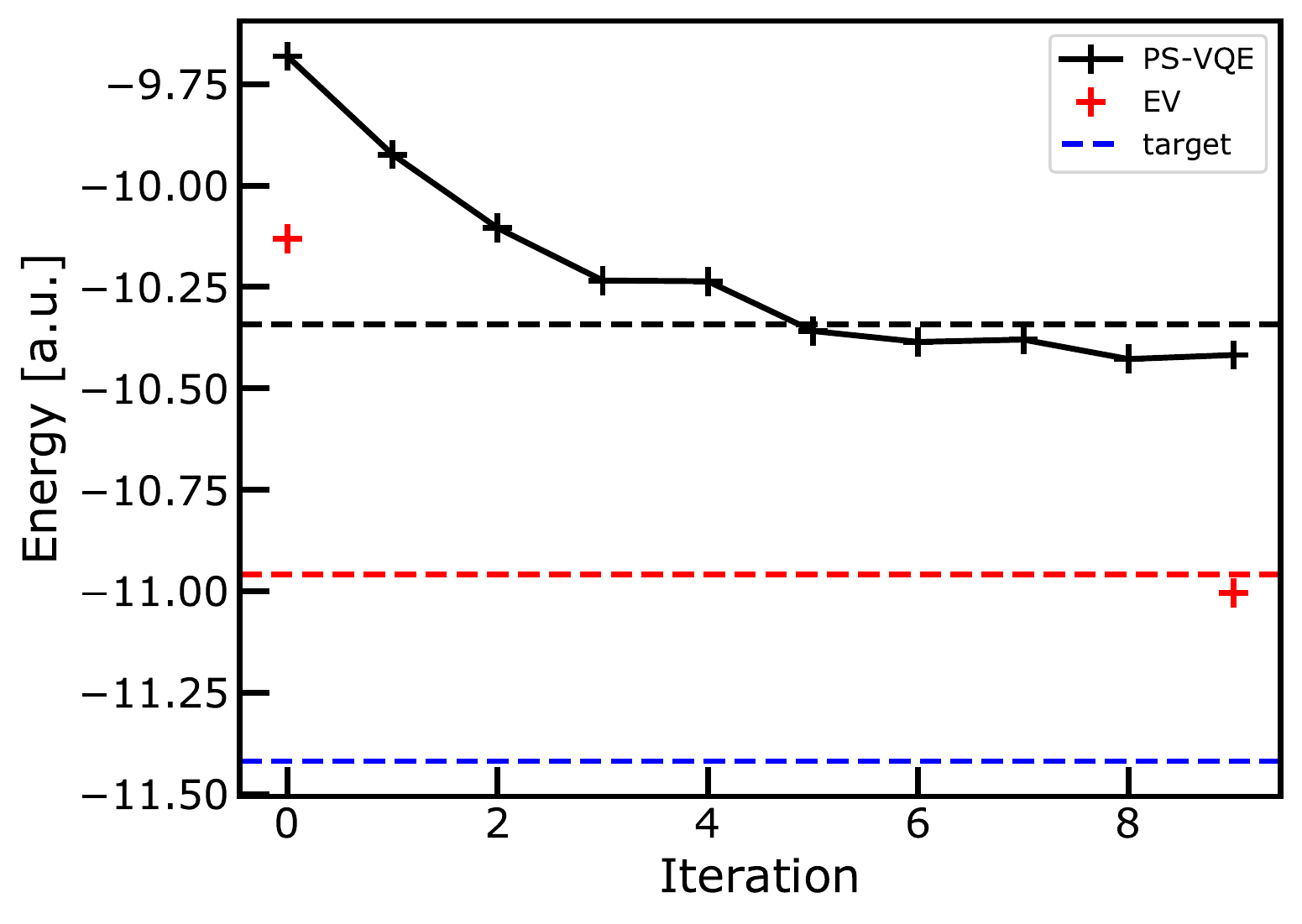}
    \caption{Optimization trace of a 6 qubit RG simulation at $g=-0.9$, targeting the blue dashed line (ground state energy in the seniority-zero space). After an energy estimation using postselection (black dashed line) and echo verification (red dashed line) at optimized parameters from a noiseless simulation, all parameters were perturbed by a random variable drawn from $[-0.25, 0.25]$, and reoptimized over $9$ iterations of Conjugate Model Gradient Descent, using the post-selected experimental data as a cost function. After perturbation and after convergence, a single call was made with the converged parameters to an echo verified energy estimation (red cross).}
    \label{fig:optimization_figure}
\end{figure}

We test the hyperparameter-tuned Conjugate Model Gradient Descent algorithm on the problem of optimizing the UpCCD ansatz for the ground state of the RG Hamiltonian at a coupling $g=-0.9$.
In order to demonstrate convergence, we perturb our ansatz parameters from values optimized on a noiseless classical simulation (using the L-BFGS-B optimizer as described above), by a random variable drawn from $[-0.25, 0.25]$.
From this point, Conjugate Model Gradient Descent converged in $9$ iterations with each iteration requiring $46$ calls to the cost function ($414$ calls total). 
We observe that the optimizer successfully finds an energy below the initial point ($0.07 a.u.$), which demonstrates the well-known VQE ability to mitigate coherent noise~\cite{Peruzzo2014Variational, McClean2015Theory}.
This improvement is reflected in the results mitigated with echo verification as well, which demonstrated a $0.04 a.u.$ reduction in error.
However, this is a relatively small reduction compared to the overall error observed.
We note that this is a significantly higher error than that observed in Fig.~\ref{fig:RG_figure_small}, which we attribute to poor calibration on the day.
If the absolute gain in energy was replicated in Fig.~\ref{fig:RG_figure_small}, this would account for $\sim 40\%$ of the overall error.
However, the relative gain in error in this case was only $\sim 10\%$.
Ultimately, as the cost of optimization was already significant enough for this $6$-qubit example, and as the number of calls for Conjugate Model Gradient Descent scales as $O(\# \mathrm{parameters})\sim O(N^2)$, we did not see it practical to continue this line of research in this work.
With current qubit counts and shot repetition rates the optimization of variational algorithms of meaningful size remains a major obstacle.

\section{Additional results and analysis}

\subsection{Virtual distillation with and without postselection}\label{app:VD-PS-comparison}
\begin{figure}
    \centering
    \includegraphics[width=\columnwidth]{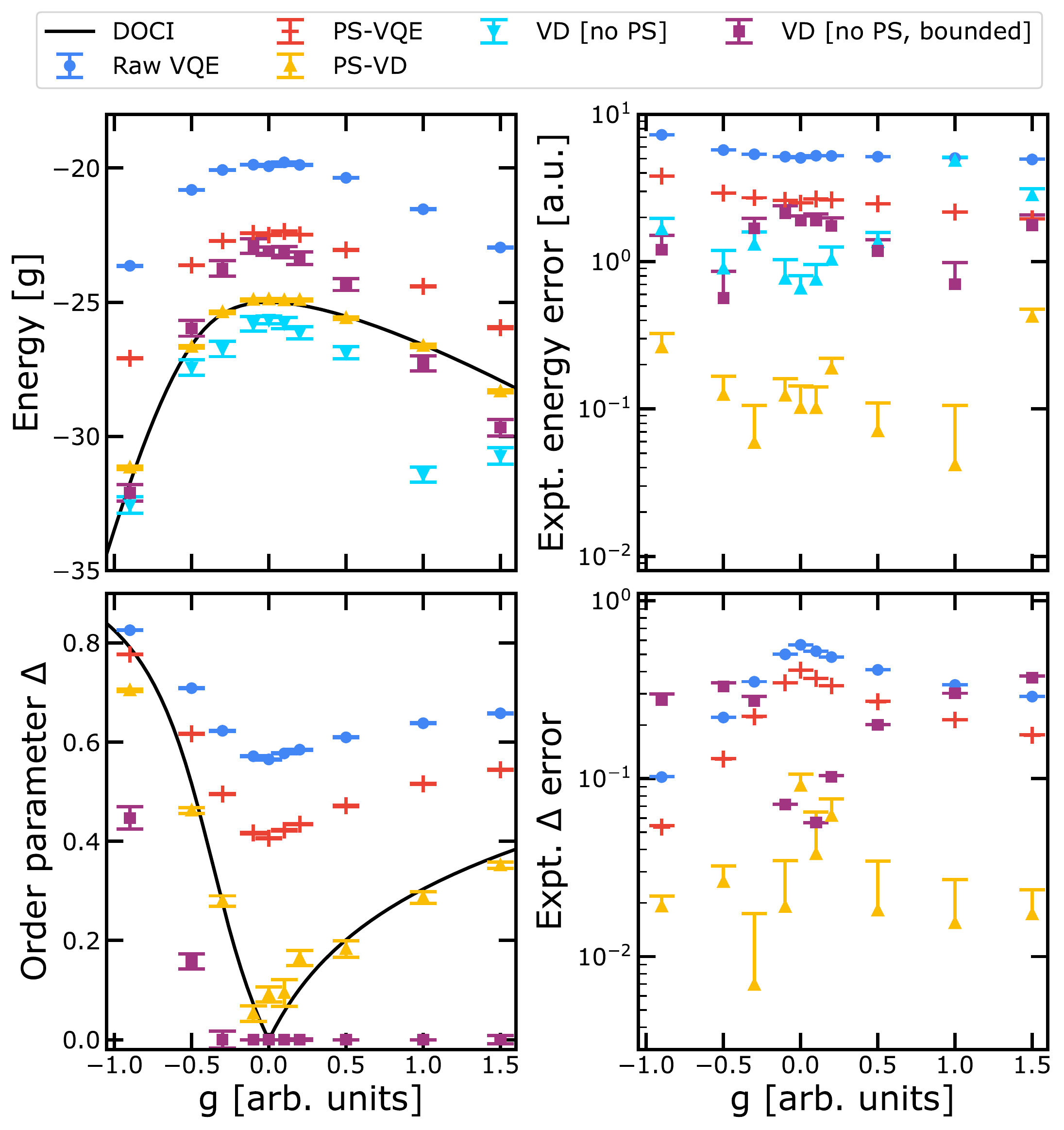}
    \caption{Comparison of virtual distillation with postselection (yellow), without postselection (light blue), and without postselection but while bounding expectation values of $Z_j$ and $D_{ij}^{\pm}$ in $[-1,1]$ (purple), for a 10-qubit simulation of ground states of the RG Hamiltonian across a range of $g$ values. Raw VQE (blue), postselected VQE (red), and post-selected VD [PS-VD] data taken from Fig.~\ref{fig:RG_figure}. VQE error bars obtained by error propagation (1 standard deviation, see App.~\ref{app:errorbars}), VD error bars obtained by bootstrapping (1 standard deviation).}
    \label{fig:VD_PS_comp}
\end{figure}

In this appendix, we show the effect that postselection has on virtual distillation.
In Fig.~\ref{fig:VD_PS_comp}, we repeat the plot of Fig.~\ref{fig:RG_figure}, but with data from VD without postselection on top.
(Other lines are removed to make the plot clearer.)
We see that by itself, VD [light blue curve] outperforms postselection in terms of energy estimation across most points of the energy curve, but it is not variationally bound.
This is because without postselection, VD returns unphysical estimates $|\langle Z_i\rangle|\geq 1$ and $|\langle D_{ij}^{\pm}\rangle|\geq 1$.
(This is a consequence of the two copies of the state not necessarily being subject to the same noise.)
As a consequence of this, our estimate of the order parameter is complex and non-physical, and therefore not plotted.
This issue can be rectified somewhat by bounding the estimates of $\langle Z_i\rangle$ and $\langle D_{ij}^{\pm}\rangle$ to lie in $[-1, 1]$.
Performing this (Fig.~\ref{fig:RG_figure}, purple curve) allows for an estimate of the order parameter to be made, however it is clearly qualitatively incorrect.
Moreover, although the energies are shifted up, some are still not variationally bounded, and those that are, often overshoot the energy, making the absolute error worse.
(The variational bound could be rectified by enforcing positivity conditions on the set of generated estimates~\cite{Rubin2018Application}.)
In summary, the combination of postselection and virtual distillation is seen here to have a greater effect than the sum of its parts for energy error.

We attribute the gain from postselection in virtual distillation to two sources.
Firstly, postselection removes final readout noise, which VD does not naturally correct against.
(As the estimation of $\mathrm{Tr}[\rho^2]$ involves a highly correlated measurement, this cannot be easily corrected by most readout correction schemes designed to estimate local Pauli expectation values.)
Secondly, though we are in principle only post-selecting on the sum of the number of excitations in $\rho^{(1)}$ and the number of excitations in $\rho^{(2)}$ being equal to $N$, when combined with virtual distillation this projects out all noise such as T1 decay and suppresses any particle-non-conserving noise to second order.
In short, this is because breaking number conservation separately on $\rho^{(1)}$ and $\rho^{(2)}$ yields states which do not overlap (as they must have different particle number), and these states cancel out when taking the product $\rho^{(1)}\rho^{(2)}$.
Let us study this in more detail.
Consider a post-selected estimate of $\mathrm{Tr}[\rho^2 O]$ using VD for $O=Z_j$ (as described in the main text).
After postselecting on $\sum_j(1\otimes Z_j + Z_j\otimes 1)=N$, we can write our global state as
\begin{equation}
    \rho_2=\sum_{p,q,r,s}c_{p,q,r,s}|p\rangle\langle q|\otimes |r\rangle\langle s|,
\end{equation}
where $p, q, r, s$ index the basis states of both systems.
(Following projection, $\rho_2$ will no longer be a product state.)
Let us write $n_p=\langle p|\sum_j Z_j|p\rangle$ etc as the number of excitations in these basis states (i.e. the Hamming weight of the index), and our projection requires $c_{p,q,r,s}=0$ unless $n_p+n_r=n_q+n_s=N$.
(This assumes WLOG we are at half-filling, and ideally $n_p=n_r=n_q=n_s=N/2$.)
Then, as $O_s$ preserves particle number for $O=Z_j$, the only contribution to
\begin{align}
    &\mathrm{Tr}\big[\rho_2 S^{(2)}(Z_j\otimes I + I\otimes Z_j)\big] \nonumber\\ &= \sum_{p,q,r,s}c_{p,q,r,s}\delta_{s,p}\delta_{r,q}\Big(\langle s|Z_j|s\rangle + \langle r|Z_j|r\rangle\Big),
\end{align}
comes from those terms $c_{p,q,p,q}$.
(The same is true for our estimate of $\mathrm{Trace}[\rho_2S^{(2)}]$.)
This implies that a non-zero contribution to $\mathrm{Trace}[\rho_2S^{(2)}]$ comes only from matrix elements $|p\rangle\langle q|\otimes |p\rangle\langle q|$ where $n_p=\frac{N}{2}+\delta$, $n_q=\frac{N}{2}-\delta$.
When $\delta=0$, our matrix element $|p\rangle\langle q|\otimes |p\rangle\langle q|$ is in the number-conserving sector.
When $\delta\neq 0$, our matrix element corresponds to a product of coherent superpositions between the $N/2+1$ and $N/2-1$ sector on both qubits.
This implies that noise channels such as the T1 channel will be completely mitigated as these off-diagonal elements do not exist.
Coherent noise may cause some of these off-diagonal elements to appear (consider e.g. a single-qubit $X$ rotation on the state $\frac{1}{\sqrt{2}}(|01\rangle+|10\rangle)$.), however this is required to happen on both states to contribute, which suppresses it to second order.
(This is in contrast to coherent noise that preserves number, to which we can be first-order sensitive.)
The above analysis has assumed that the postselection is perfect; readout noise would complicate the above.

\subsection{Distribution of errors in Pauli expectation value estimation}\label{app:pauli_error_histogram}
\begin{figure}
    \centering
    \includegraphics[width=\columnwidth]{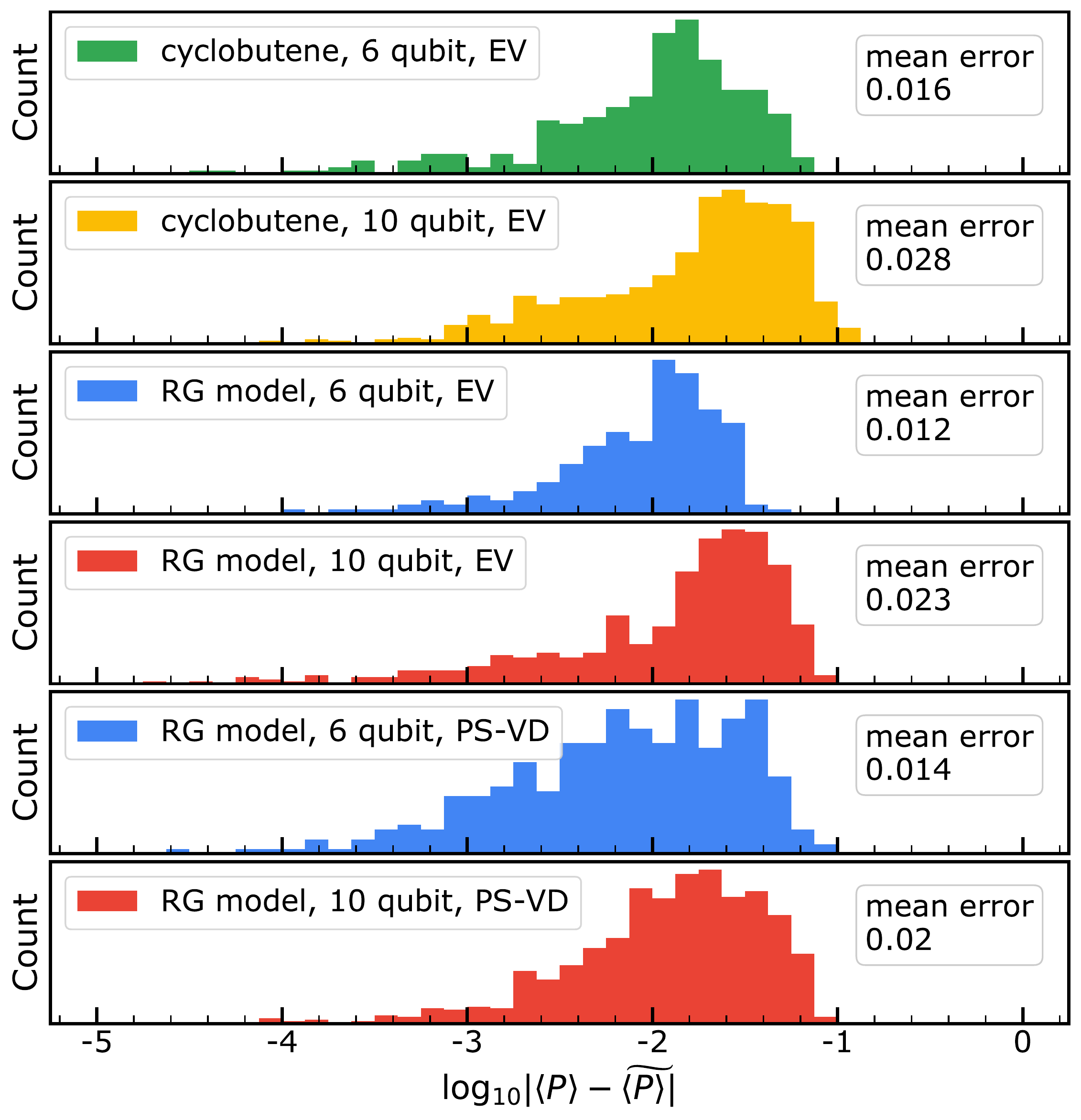}
    \caption{Histogram of the expectation value error in Pauli operators $P=Z_j$, $P=Z_iZ_j$ or $P=D_{ij}^{\pm}$ (Eq.~\eqref{eq:Dij}), across all data taken for the experiment mentioned. Mean error across the dataset is given.}
    \label{fig:pauli_error_histogram}
\end{figure}

In Fig.~\ref{fig:pauli_error_histogram}, we plot a histogram of the error in estimating the expectation values of Pauli operators, taking data from Fig.~\ref{fig:RG_figure}, Fig.~\ref{fig:RG_figure_small}, and Fig.~\ref{fig:cyclobutene_fig}.
We observe that the mean error in both cases for cyclobutene is slightly worse than for the RG model, but only by a small percentage.
(This justifies our claim in Sec.~\ref{sec:scaling} that a $0.1~\mathrm{a.u.}$ error in the RG Hamiltonian is approximately $1-2$ orders of magnitude larger than chemical accuracy for a $10$-qubit system.)
As the difference between systems here is minimal (changing only the value of some virtual Z rotations), we can only attribute this difference to the performance of the device whilst taking these datasets.
Going from $6$ to $10$ qubits increases the mean error in all experiments by a factor $1.5-2\times$.
As the Hamiltonian for both the RG model and cyclobutene has a number of terms scaling as $O(N^2)$; assuming that the errors follow a roughly Gaussian distribution would predict the energy error scales $O(N)\times$ the error per each individual operator.
This then predicts a gain in energy of $2.5-3.3\times$, which lies in between the observation of Fig.~\ref{fig:cyclobutene_fig} and Fig.\ref{fig:scaling_figure}.
This suggests that the RG model energy estimation at large $N$ may have had some benevolent cancellation of noise, and likewise for the cyclobutene energy estimation at small $N$.

\subsection{Smaller studies of the RG Hamiltonian}\label{app:RG_smaller}
\begin{figure*}
    \centering
    \includegraphics[width=\columnwidth]{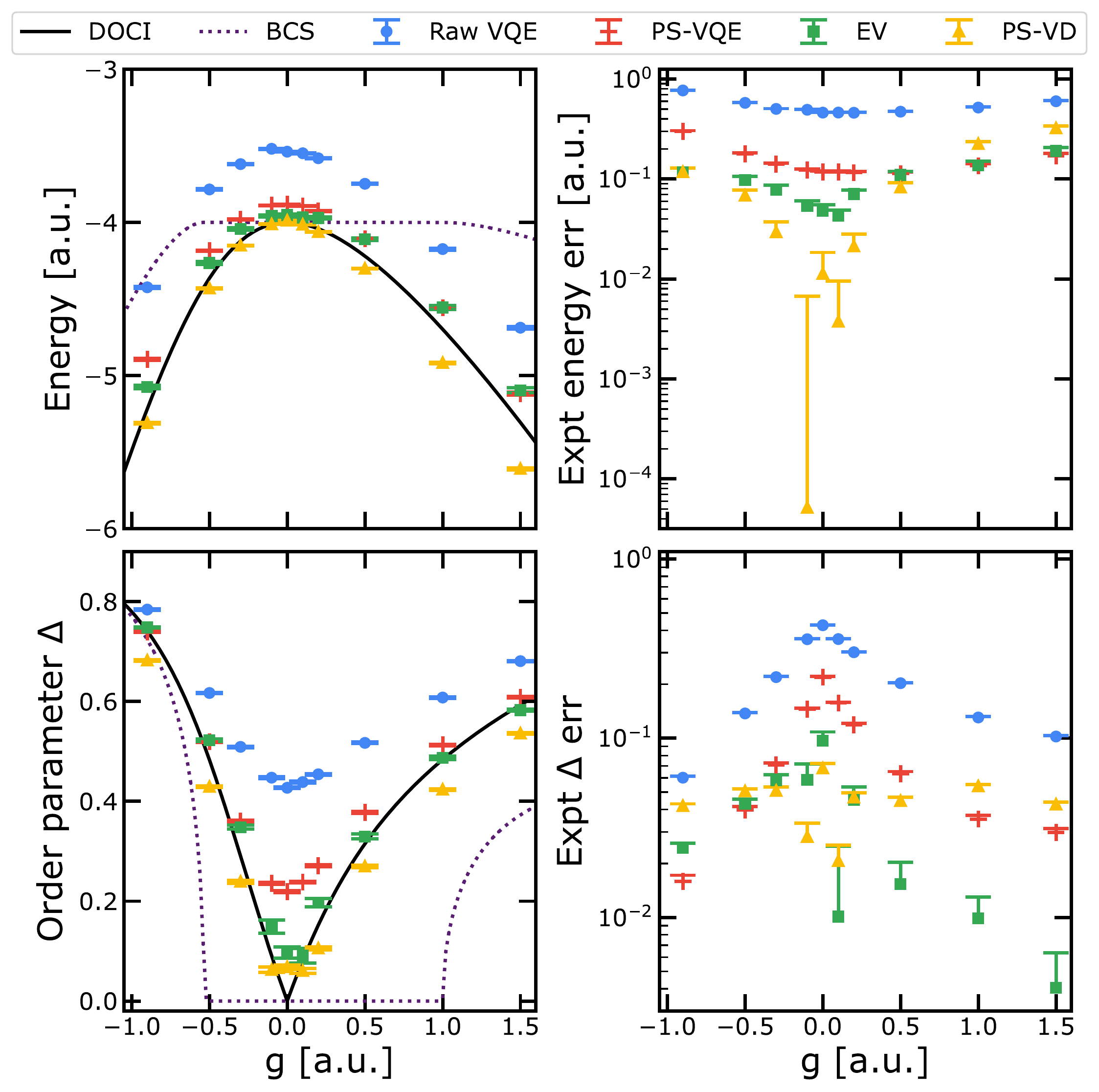}
    \includegraphics[width=\columnwidth]{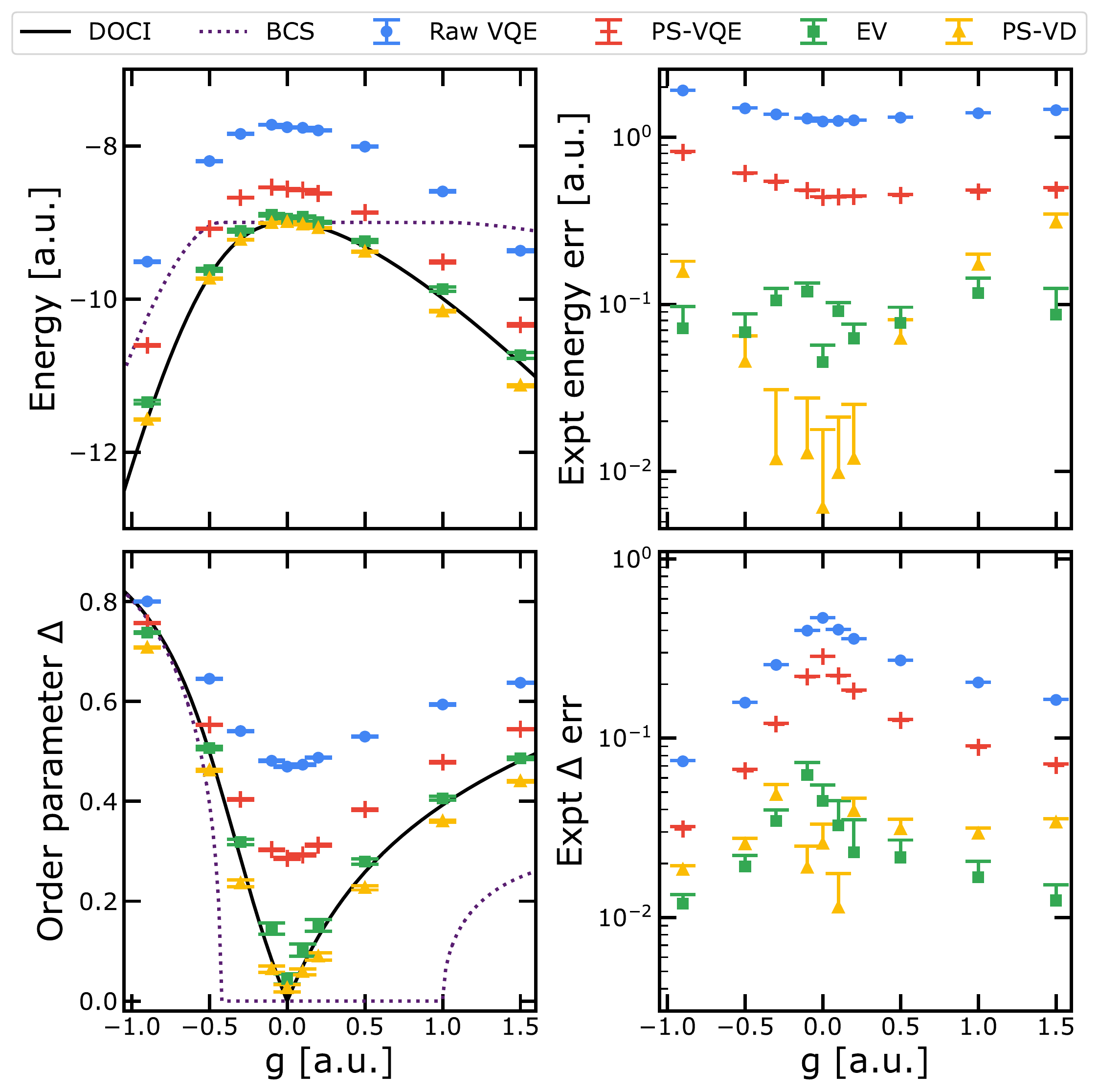}
    \includegraphics[width=\columnwidth]{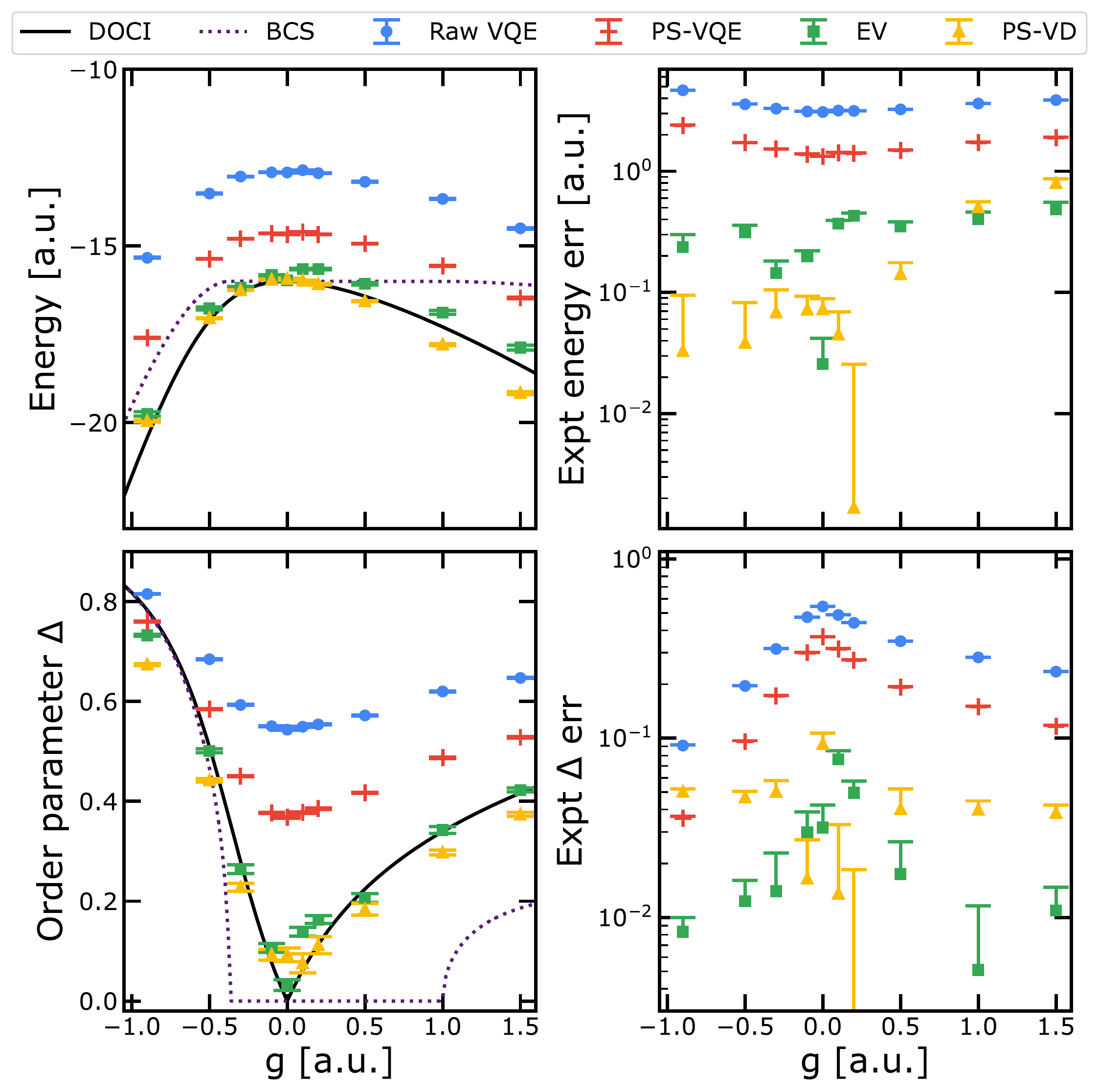}
    \caption{Identical experiment to Fig.~\ref{fig:RG_figure}, but for 4 (top-left), 6(top-right), and 8 (bottom) qubits instead of 10. Aggregate data is used in Fig.~\ref{fig:scaling_figure}.
    (top-left) Energy plot for the RG system as a function of the coupling parameter $g$, for an unmitigated state preparation [blue circles], and state preparation mitigated by postselection [red crosses], echo verification [yellow triangles], and postselected virtual distillation [green squares]. This is compared to the exact DOCI result [black solid line], and BCS [purple dashed line]. (top-right) Log plot of experimental energy error (ignoring the model error from the UpCCD approximation). (bottom-left) Superconducting order parameter for the RG Hamiltonian ($\Delta = \frac{1}{N}\sum_{j,\sigma}\sqrt{\langle n_{j\sigma}^{2}\rangle-\langle n_{j\sigma}\rangle^2}$), again compared to classical models. (bottom-right) Experimental error in estimating the superconducting order parameter vs the target state within the UpCCD approximation (again ignoring model error). 1 std. dev. error bars estimated by either propagating variance (Raw VQE, PS-VQE) or bootstrapping (EV, PS-VD).}
    \label{fig:RG_figure_small}
\end{figure*}

In Fig.~\ref{fig:RG_figure_small}, we present experimental simulations of the ground state of the RG Hamiltonian for $4$, $6$, and $8$ qubits.
The method to produce these figures is identical to that used in the production of Fig.~\ref{fig:RG_figure}, save for the number of qubits and shots used.
Aggregated data from these figures was used to generate the scaling plots of Fig.~\ref{fig:scaling_figure}.

\end{document}